\documentclass[aps,pre,twocolumn,showpacs]{revtex4} 
\let\revappendix\appendix

\usepackage{amsmath} 
\usepackage{amssymb} 
\usepackage{graphicx} 
\usepackage{graphics}
\usepackage{epsfig}
\usepackage{multirow}
\usepackage{color}
\usepackage{slashed}
\usepackage{etoolbox} 
\newcommand\nn{\nonumber}
\newcommand\ba{\begin{eqnarray}}
\newcommand\ea{\end{eqnarray}}
\newcommand\be{\begin{equation}}
\newcommand\ee{\end{equation}}

\begin{document}
\title{Form factor ratio from unpolarized elastic electron proton scattering}
\author{Simone Pacetti}
\affiliation{\it  Dipartimento di Fisica e Geologia, and INFN Sezione di Perugia, 06123 Perugia, Italy}
\author{Egle~Tomasi-Gustafsson}
\email{etomasi@cea.fr}
\affiliation{\it IRFU, CEA, Universit\'e Paris-Saclay, 91191 Gif-sur-Yvette Cedex, France}
\date{\today}
\begin{abstract}
A reanalysis of unpolarized electron-proton elastic scattering data is done in terms of the electric to magnetic form factor squared ratio. This observable is in principle more robust against experimental correlations and global normalizations. The present analysis shows indeed that it is a useful quantity that contains reliable and coherent information. The comparison with the  ratio extracted from the measurement of the longitudinal to transverse polarization of the recoil proton in polarized electron-proton scattering shows that the results are compatible within the experimental errors. Limits are set on the kinematics where the physical information on the form factors can be safely extracted. The results presented in this work bring a decisive piece of information to the controversy on the deviation of the proton form factors from the dipole dependence.
\end{abstract}
\maketitle
Hadron form factors (FFs) contain essential information on the electric and magnetic charge currents in the hadron and constitute a very convenient parametrization that enters in theoretical models and description of experimental observables concerning the three-leg vertex proton-proton-photon.
It has been assumed for long time that the proton electric FF, as well as the magnetic FFs of the proton and neutron, normalized to their magnetic moment, have a $Q^2$-dipole dependence ($Q$ is the four momentum of the photon): 
\ba
G_D(Q^2)=\left(1+\frac{Q^2 {\rm [GeV]}^2}{0.71}\right)^{-2}\,, \nn
\ea
whereas the neutron electric FF is essentially zero. Evidence for the deviation of the proton electric FF from the dipole form was  present since the 70's, in experiments as well as in theory. However, the dipole parametrization was commonly accepted due to the following facts:
\begin{itemize}
\item From the classical point of view, in the non-relativistic approximation, FFs are Fourier transforms of the spatial densities of electric charge and magnetization of the nucleon; the dipole approximation corresponds to an exponential distribution and the parameter 0.71~GeV$^2$ corresponds to a quite reasonable root mean square radius of the proton of 0.81 fm.
\item From the QCD point of view, as elastic FFs represent the probability that a proton remains in its ground state after that each of its valence quarks has received a momentum squared, $Q^2$, transferred by the virtual photon, scaling laws predict a $(1/Q^2)^2$ dependence of the amplitude of the process~\cite{Matveev:1973ra,Brodsky:1973kr} (corresponding to two gluon exchange, the minimum number of exchanges needed for sharing the momentum among the three valence quarks). 
\end{itemize}
The {\it reduced} cross section of  electron-proton elastic scattering, in Born approximation, i.e., by considering only one-photon exchange,  $\sigma_{\rm red}$,  is linear in the variable $\epsilon=[1+2(1+\tau)\tan^2(\theta_e/2)]^{-1}$, being $\theta_e$ the electron scattering angle in the proton rest frame, and $\tau=Q^2/(4M^2)$, and it reads
\ba
\sigma_{\rm red}
(\theta_e,Q^2)&=&
\nn \\
&&
\hspace*{-2truecm}
\left [1+2\displaystyle\frac{E}{M}\sin^2(\theta_e/2)\right ]\displaystyle\frac
{4 E^2\sin^4(\theta_e/2)}{\alpha^2\cos^2(\theta_e/2)}\times
\epsilon(1+\tau)\displaystyle\frac{d\sigma}{d\Omega}\nn \\
&=&\epsilon\, G_E^2+\tau \,G_M^2\,, 
\label{eq:sigma}
\ea
where $M$ is the proton mass, $E$ and $d\sigma/d\Omega$ are the electron initial energy and the differential cross section in the proton rest frame, and $G_E$ and $G_M$ are the proton electric and magnetic Sachs FFs. The measurement of the differential (reduced)  cross section at fixed $Q^2$, for
different angles, allows to extract the squared values of the 
FFs, $G_E^2$ and $G_M^2$, as the slope and the intercept (multiplied by $\tau$),
respectively, of this linear
distribution (Rosenbluth separation~\cite{Rosenbluth:1950yq}).

However, hints from experiments and theory cast some doubts on the accuracy of the dipole approximation. 
\begin{itemize}
\item \underline{In experiment:}\\
Series of measurements at moderate $Q^2$ at DESY~\cite{Bartel:1973rf} (single arm), at Cambridge electron accelerator~\cite{PhysRevD.4.45} with $ep$ coincidence and from proton and quasi-elastic deuteron scattering~\cite{PhysRevD.8.753},  found a decrease of the electric FF with $Q^2$, with respect to the dipole, in the limit of the errors. In these experiments radiative corrections were either measured, or controlled through a comparison between proton 
and deuteron 
targets, and never exceeded 20\%. Further measurements at Mainz~\cite{Borkowski1975461} as well as the SLAC experiment~\cite{PhysRevLett.20.292} showed also a deviation of the ratio $\mu G_E/G_M$ from unity, being $\mu$ the proton magnetic moment in units of the Bohr magneton. 
The aim to learn about nucleon structure till the highest transferred momenta, justified the extraction of $G_M$ under the hypothesis $G_E=0$ or $G_E=G_M/\mu$~\cite{Litt197040}, as in any case, the electric contribution to the unpolarized cross section is suppressed by that factor of $\tau$ with respect to the magnetic term.
At larger $Q^2$ the large errors and other type of corrections, in particular radiative corrections, were applied following commonly accepted ansatzs and, in our opinion,  not critically revised. Moreover, recent, dedicated measurements at JLab~\cite{Qattan:2004ht}, as well as reanalyses of existing data~\cite{Christy:2004rc} confirmed the scaling FFs behavior: $G_E\simeq G_M/\mu$.

\item \underline{ In phenomenology:}\\ 
Typically the electric and magnetic distributions do not have to follow a priori similar $Q^2$ behavior. Moreover, magnetic and electric FFs of neutron and proton may be different, as they do not have the same quark content. Some models predicted the decrease of the electric FF long before the data appeared, as the two component model of Ref.~\cite{Iachello:1972nu}, built on vector meson dominance, or  the soliton model~\cite{Holzwarth:1996xq} that attributed the decrease of the electric FF to approximations of relativistic effects, or the di-quark  model of Ref.~\cite{Kroll:1967it}.
\end{itemize}


The doubts on the deviation from the dipoIe became evident with the advent of polarization experiments. In the 70's Akhiezer and Rekalo~\cite{Akhiezer:1968ek,Akhiezer:1974em} show that the polarization of the scattered proton in the scattering of longitudinally polarized electrons on an unpolarized target (or the asymmetry in the scattering of longitudinally polarized electrons on a transversely polarized target) contains an interference term proportional to the product $G_E\,G_M$. This observable would therefore be more sensitive to a small electric contribution, and even to its sign (particularly important for the neutron case). The suggested polarization method could be realized only in the years 2000, following the advent
of high-intensity electron beams, large acceptance detectors 
and due to the tremendous progress in polarization techniques for beams,
targets and hadron polarimetry. A measurement of the ratio of transverse and longitudinal polarization of the recoil proton gives a direct measurement of the ratio of electric and magnetic FFs, $R=G_E/G_M$: 
\ba
\displaystyle\frac{P_t}{P_\ell}= - 2\cot(\theta_e/2) \displaystyle\frac{M_p}{E+E'}\displaystyle\frac{G_E}{G_M}\,,
\label{eq:final}
\ea
and is free from systematic errors coming from the beam polarization and the analyzing powers of the polarimeter.

The data based on the Akhiezer-Rekalo method, mostly taken by the JLab GEp collaboration (\cite{Puckett:2011xg} and references therein),  showed with unprecedented precision that the ratio of electric to magnetic FFs decreases as $Q^2$ increases.  


However, injecting the polarization ratio into the unpolarized cross section would modify $G_M$ by $3\%$ at most, within the experimental errors~\cite{Brash:2001qq}; showing that problem is not at the level of the observables (i.e., unpolarized cross section and ratio of longitudinal to transverse proton polarization), within the experimental errors. 

Different conjectures or possible solutions to this problem were discussed in the literature. 

The experimental data were generally corrected by first order  radiative corrections, following the calculation for example, of Ref.~\cite{Mo:1968cg}.  This calculation contained approximations that were justified at small $Q^2$ and/or small acceptance detectors, but that may not hold at large incident and scattered electron energy and momentum, as well as at large momentum transfer. Radiative corrections become huge at large $Q^2$ especially for the unpolarized cross section. They are as large for the polarized cross sections, but mostly cancel in the polarization ratio. 

A revision removing some of these approximations, was published~\cite{Maximon:2000hm}, whereas a critical analysis of first order calculations is available  in Refs. \cite{Gramolin:2016hjt,Gerasimov:2015aoa}. 
New results on radiative corrections based on the lepton structure functions method, that takes into account higher orders in the leading logarithm approximation \cite{Kuraev:1985hb}, can bring FF results in agreement~\cite{Bystritskiy:2007hw}. This was confirmed later on by a more extended calculation, including hard photon emission~\cite{Kuraev:2013dra}.  Note that, since the polarization data were made available, no experiment was performed to verify the kinematics and the radiative emission with a precise measurement of the four momenta of both outgoing particles and/or radiated particles in $ep$ scattering. This would allow to revise critically the assumptions and corrections applied to the previous (unpolarized) measurements.

The discrepancy was attributed by several authors to the presence of a two-photon exchange contribution. From that time, a lot of theoretical and experimental work was devoted to this subject. Note that the two photon contribution was already discussed in the literature in the 70's~\cite{Boitsov:1972if,Lev:1975,Franco:1973uq} and recently rediscussed:  for $ed$ elastic scattering \cite{Rekalo:1999mt}, and $ep$ elastic scattering \cite{Guichon:2003qm}. A series of articles on model independent properties of the two-photon contribution on different processes:
$ep$ scattering~\cite{Rekalo:2003xa, Rekalo:2003km,Rekalo:2004wa},
$\bar p p \to e^+e^-$~\cite{Gakh:2005wa}, $ e^+e^-\to \bar p p$~\cite{Gakh:2005hh,Chen:2008hka,Zhou:2011yz},
$e^-  He^4$ scattering and $e^+ e^-\to \pi^+ \pi^-$~\cite{Gakh:2008fb}, showed clearly the consequences of the non-applicability of the one-photon approximation, making necessary a serious revision of most of the obtained results. As an example, two-photon contributions would induce non-linearities in the Rosenbluth fit as the hadronic current would be parametrized  by three structure functions, of complex nature and depending on two  kinematical variables, instead than by two real FFs functions of $Q^2$. The extraction of the real FFs would still be possible, but requiring: either polarized electron and positron beams, applying the Akhiezer-Rekalo method to the sum of the cross sections, where odd terms disappear; or measuring five T-even or three T-odd polarization observables, including triple spin observables, which appears very difficult.

Reanalyses of $e^+p/e^-p$ data~\cite{Arrington:2007ux,TomasiGustafsson:2009pw,Alberico:2009yp}, searching for non-linearities of the reduced cross section,  gave  negative results, the slope of the Rosenbluth plot being driven by $Q^2$ and $\epsilon$-dependent radiative corrections (the  slope of the uncorrected cross section becoming even negative for Q$^2~>~2$~GeV$^2$~\cite{TomasiGustafsson:2004ms});

Model calculations of the hadronic two-photon exchange contribution were developed, giving quantitatively different results since the physical reasons for an enhancement of this term beyond the $\alpha$-counting expectation differ essentially from a model to another~\cite{PhysRevD.72.013008,Borisyuk:2008es,Kivel:2009eg}. Several measurements were proposed~\cite{Gramolin:2011tr,Milner:2012zz,Bennett:2012zza} and the results show that  an asymmetry between electron and positron scattering exists indeed, and may reach 6-7\%. However, most of the asymmetry comes from the interference between initial and final photon emission, and it is highly reduced when  the data are properly radiatively corrected. The size of the additional two-photon contribution does not exceed  the expected size from $\alpha$ counting (2-3 \%); moreover, the measurements, being performed at low $Q^2$, do not show  evident increase with $Q^2$. Note that an effect growing with $Q^2$ and reaching $6\%$ at $Q^2\simeq 6$~GeV$^2$ is necessary to bring in agreement the data on the ratio $G_E/G_M$ extracted from the  Akhiezer-Rekalo and the Rosenbluth methods.

\section{Reanalysis of existing data}
Problems of parameter correlations and limits inherent to the Rosenbluth have been discussed in Ref.~\cite{TomasiGustafsson:2006pa}. Previous global analysis were done, discussing in particular the problem of normalization among different sets of data, the omission of some of the data points \cite {Arrington:2003df}, reconsidering radiative corrections \cite{Gramolin:2016hjt}.

Here we suggest the following procedure to extract the FFs information from the unpolarized cross section. Instead of  extracting separately $G_E$ and $G_M$, we write the reduced cross section given in Eq.~(\ref{eq:sigma}) as
\be 
\sigma_{\rm red}= G_M^2(R^2 \epsilon +\tau)\,, 
\label{eq:sred}
\ee
where $G_M^2$ and $R^2=(G_E/G_M)^2 $ are considered as independent parameters. The unpolarized data are fitted at fixed $Q^2$. The procedure has the advantage to extract directly the ratio, by automatically accounting for the effect of the correlations between $G_E$ and $G_M$. The parameter $R^2$ represents directly the deviation of the linear dependence of the cross section from a constant in $\epsilon$, whereas general normalization and systematic errors would be absorbed by $G_M^2$. 

The results of the fit on the considered measurements of the unpolarized elastic $ep$ cross section are summarized in Appendix \ref{app:fit}. 

If, for some of the data, we recover values and errors consistent with the original publication, the data from Ref.~\cite{Andivahis:1994rq} deserve a specific discussion. This work is especially  representative, as it extends the individual FF extraction by the Rosenbluth method to the largest values of $Q^2$. 

\subsection{Data from Ref.~\cite{Andivahis:1994rq}}
The original cross section data are reported in Appendix \ref{app:andi}, Table~\ref{tab:andi}, and  the individual fits, at each $Q^2$ are illustrated in Fig.~\ref{fig:andi}. The data of Ref.~\cite{Andivahis:1994rq}, with eight points and  two settings, span the region $1.75 \le  Q^2 \le 8.83$~GeV$^2$.  The two settings will be indicate as  high energy (HE) and low energy (LE) experiments. 

\subsubsection{Analysis I}
In the original paper the measured cross sections were published, warning that an uncertainty of $\pm 5\%$ affected the second setting, due to a poor knowledge of the acceptance of the spectrometer. This error, however, was not added to the tabulated error. Instead,  it was taken into account as a constant relative correction, according to the following procedure:
\begin{itemize}
\item 
For the two lowest values $Q^2 =1.75$ and $2.50 $ GeV$^2$, the cross section was measured at each setting at the lowest $\epsilon$, and showed a larger value from the LE setting by 4-5\%.
 \item Assuming a linear $\epsilon$ dependence of the reduced cross section, i.e., the dominance of the one-photon exchange mechanism, a fit of the HE data was done and the LE energy point was renormalized to lie on the straight line. 
 \item  The same constant normalization $C=0.956$, fixed on the low $Q^2$ point,  was applied to the cross section at all $Q^2$.
 \end{itemize}
This procedure has the effect to enhance the slope, increasing the FF ratio. Note that for $Q^2=6$ and 7~GeV$^2$ only two points are present. The renormalization (lowering) of the first point changes essentially the slope of the linear fit.
\subsubsection {Analysis II}
We  recalculate the ratio using the data as published, without renormalizing the two settings and considering the LE points as additional, independent measurements.  In this case  the data points at $Q^2 =1.75 $ and $2.5 $~GeV$^2$ are both included in the fit, constraining the fit to an average value.
\subsubsection{Analysis III}
We fit only the HE points. (excluding therefore the two points at  $Q^2=6$ and 7~GeV$^2$). We find a slope consistent with analysis II, although affected by larger errors, as the number of points is smaller.  
\subsubsection {Analysis IV}
\begin{figure}[h!]
\begin{center}
\includegraphics[width=\columnwidth]{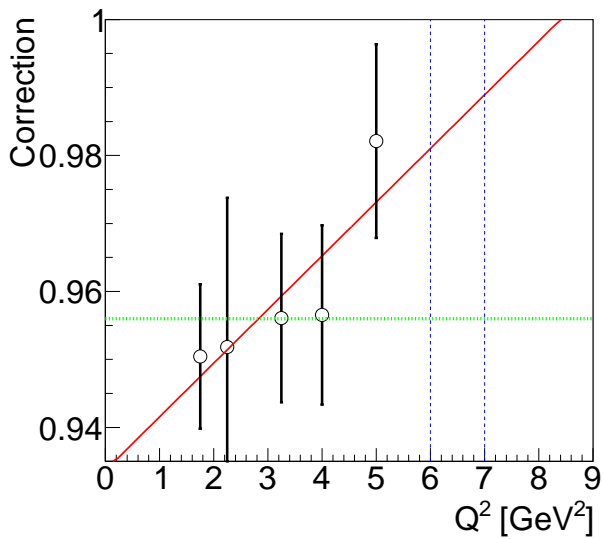}
\caption{Color online. Correction factor as a function of $Q^2$. A linear fit (red line) shows an increasing of the factor. The dashed (blue) lines indicate that the extrapolated correction for the two HE points would be close to $1\%$ instead than $\simeq 5\%$, as applied in the original paper.}
\label{Fig:correction}
\end{center}
\end{figure}

We repeat the normalization procedure, by aligning the LE point on the straight line fitting the HE points. We note a systematic increase of the normalization factor  (Fig. \ref{Fig:correction}, and Table \ref{table1}).  It is well known that the acceptance of a spectrometer depends on the kinematics of the particle, it is not surprising that the needed corrections decrease at large energies ($C \to 1$). Applying a normalization coefficient that is not constant with $Q^2$ but derived in order to align the LE point to the straight line defined by the HE points turns out to be equivalent to Analysis III in terms of slope and intercept. This explains the agreement between analysis III and IV . 

\begin{figure}[h!]
\begin{center}
\includegraphics[width=\columnwidth]{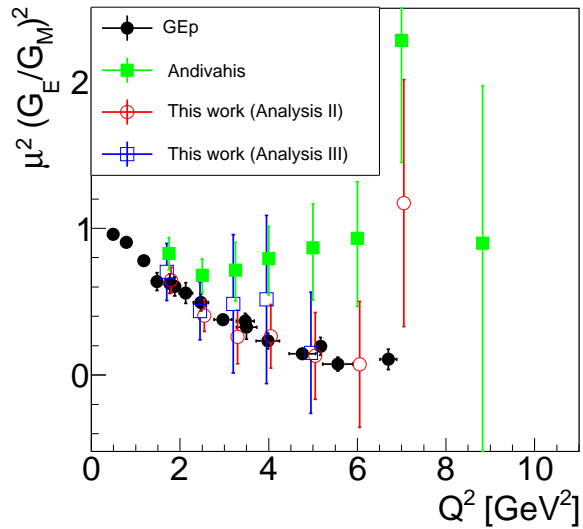}
\caption{Color online. $\mu^2 R^2= \mu^2 (G_E/G_M)^2$  as a function of $Q^2$ from (Andivahis: \cite{Andivahis:1994rq}, green solid squares) as originally published; from Analysis II: without renormalization (red open circles); from Analysis III: omitting the lowest $\epsilon$ point (blue open  squares)  compared to the values from polarization experiments (GEp: \cite{Puckett:2011xg}, black solid circles). 
 }
\label{Fig:W0}
\end{center}
\end{figure}

The results are reported in Fig.~\ref{Fig:W0} and compared to the ratio from polarization data. We may conclude that the results from Analysis II, III, and IV are consistent with the decreasing of the ratio indicated by the polarization data. Therefore a revision of the normalization factor brings the data into agreement. Moreover, at the light of all above, it is nonsense to use the FFs data from Ref.~\cite{Andivahis:1994rq} to probe the two-photon effect, as they were extracted under the hypothesis of linearity of the reduced cross section, i.e., correcting the first point  to be aligned. The results showed consistency with the hypothesis $\mu^2 R^2\simeq 1$ at large $Q^2$, as expected at that time. The tendency of the first two points to deviate from unity was operatively corrected by the renormalization procedure.

\section{Analysis and discussion of available data}

A complete discussion and data basis of unpolarized and polarized measurements can be found in Ref.~\cite{Pacetti:2015iqa}.  There, it was already noted that some unpolarized data, where radiative corrections were lower than $20\%$, indeed showed a deviation of the ratio $\mu^2 R^2$ from unity consistently with the polarization data. 

We consider below only the cross section data where the individual determination of FFs was done and the ratio was extracted from a Rosenbluth separation. The main set of data, considered in this analysis, is the one collected in Ref.~\cite{Pacetti:2015iqa}, with a  focus on  the region $Q^2\ge 1$~GeV$^2$, which includes 64 data points.
\begin{table}[h!]
\begin{center}
\begin{tabular}{|c|c|}
\hline\hline
 $ Q^2$ (GeV$^2$)& Correction \\
\hline\hline
1.75$^*$ &  0.951144 $\pm$ 0.0156952 \\
1.75         &  0.950432 $\pm$  0.0106094 \\
2.25$^*$  &  0.955992 $\pm$  0.0259077 \\
2.25         &  0.951849 $\pm$  0.0219368 \\
3.25         &  0.956075 $\pm$  0.0123809  \\
4              &  0.956552 $\pm$  0.0131748 \\
5              &  0.982138 $\pm$  0.0142443 \\                         
 \hline \hline
\end{tabular}
\caption{Normalization factor for the LE point derived from a linear fit of the HE points. The data are those of Ref.~\protect\cite{Andivahis:1994rq}. The numbers with the superscript "$*$" are directly derived from the ratio of the measured cross sections. }
\label{table1}
\end{center}
\end{table}\\
Data obtained from unpolarized  $ep$ elastic scattering are reported from Refs.~\cite{PhysRev.142.922,Bartel:1973rf,Berger197187,Christy:2004rc}, showing a  squared ratio consistent with the polarization data, in the limit of the (large) errors, are reported in Fig.~\ref{Fig:W}. The corresponding data are reported in Appendices \ref{app:jans}, \ref{app:bart},  \ref{app:berg}, \ref{app:chri}, the cross sections in Tables~\ref{tab:jans}, \ref{tab:bart}, \ref{tab:berg}, \ref{tab:chri}, and  the individual fits, at each $Q^2$ are illustrated in Figs.~\ref{fig:jans}, \ref{fig:bart}, \ref{fig:berg}, \ref{fig:chri}, respectively.

\begin{figure}[h!]
\begin{center}
\includegraphics[width=\columnwidth]{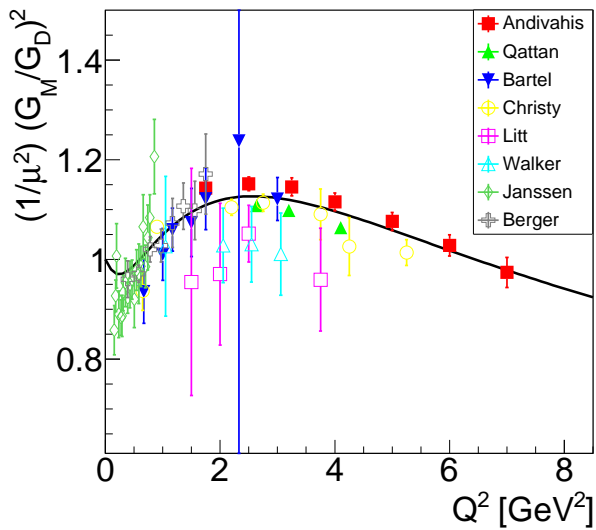}
\caption{Color online. Magnetic FF (normalized to $\mu$ and to the dipole) square as a function of $Q^2$.  Data are from  (Bartel:  \cite{Bartel:1973rf}, blue triangles down), (Christy: \cite{Christy:2004rc}, yellow open circles), (Janssen: \cite{PhysRev.142.922}, green open lozenges) (Berger: \cite{Berger197187}, grey open crosses), (Qattan: \cite{Qattan:2004ht}, green solid  triangles),  (Walker: \cite{Walker:1993vj}, cyan open triangles) and (Litt: \cite{Litt197040}, pink open squares), compared to the calculation of Ref.~\cite{Bijker:2004yu}, chosen as an example (black solid line).}
\label{Fig:GM2}
\end{center}
\end{figure}

The other parameter of the fit,  $G_M^2$, normalized to the dipole and to the proton magnetic moments (squared), is shown in Fig.~\ref{Fig:GM2}. These results are important for consistency check, in order to corroborate the suggested procedure. 
\begin{figure}[h!]
\begin{center}
\includegraphics[width=\columnwidth]{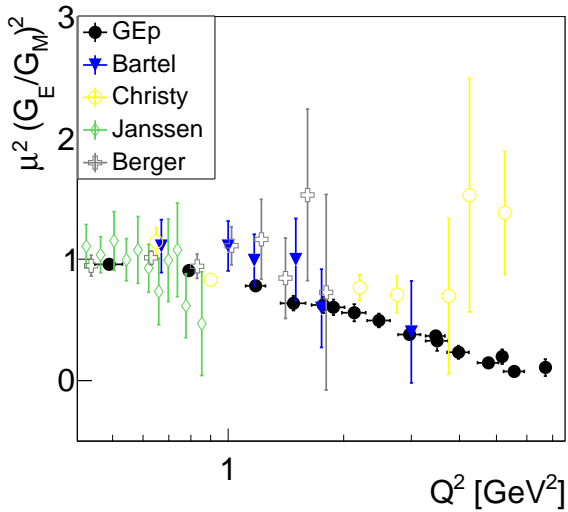}
\caption{Color online. $\mu^2 R^2= \mu^2 (G_E/G_M)^2$   as a function of $Q^2$ (Bartel:  \cite{Bartel:1973rf}, blue triangles down),  (Christy: \cite{Christy:2004rc}, yellow open circles), (Janssen: \cite{PhysRev.142.922}, green open lozenges) and (Berger: \cite{Berger197187}, grey open crosses) compared to the values from polarization experiments (GEp:~\cite{Puckett:2011xg}, black solid circles).}
\label{Fig:W}
\end{center}
\end{figure}\\
They are compared to the vector meson dominance model of Ref.~\cite{Bijker:2004yu} (black solid line), chosen as an example. As expected, the magnetic FF is better determined by the Rosenbluth fit, and the present values are consistent with the model that represents, in fact, a global fit to the data.  The numerical values  are reported in Appendix~\ref{app:fit} and Table~\ref{tab:conf1}. 
Among the available data, three sets~\cite{Walker:1993vj,Litt197040,Qattan:2004ht} show a particular behavior, that is not consistent with the previous finding, giving a value of the ratio exceeding unity and growing with $Q^2$. 
The numbers are reported in Appendices~\ref{app:walk}, \ref{app:litt},  \ref{app:qatt}, the cross sections in Tables~\ref{tab:walk}, \ref{tab:litt}, \ref{tab:qatt}, and  the individual fits, at each $Q^2$ are illustrated in Figs.~\ref{fig:walk}, \ref{fig:litt}, \ref{fig:qatt}, respectively.

For these experiments it was noted in Ref.~\cite{TomasiGustafsson:2006pa} that radiative corrections and/or correlations are especially large. The data from Ref.~\cite{Qattan:2004ht} were extracted detecting the proton instead of the electron. Besides the above mentioned corrections, at large $Q^2$ the contamination of the elastic peak by the inelastic $e+p\to e+p+\pi^0$ reaction has to be carefully subtracted~\cite{Puckett:2011xg}. 

For Refs.~\cite{Walker:1993vj,Litt197040}, $G_M^2$ extracted from the present analysis is systematically lower,  showing that these measurements may be affected by a global systematic error probably due to normalization issues, whereas the results of Ref.~\cite{Qattan:2004ht} agree with the  standard parametrization of the magnetic contribution.

Note that in Ref.~\cite{Litt197040}  a somehow arbitrary renormalization was done by "changing the normalization of the small angle data from SLAC or DESY by $\pm 1.5\%$ with respect to the large angle data (Bonn)". This normalization increased the FFs ratio towards unity.

The complete set of results in form of tables is given in the Appendixes. Concerning, in general, the elastic $ep$ cross section, several early experiments pointed out a deviation of the elastic cross section from the $(1/Q^2)^2$ behavior.  Quoting a  presentation of the data at the highest available transferred momenta, from Nobel prize R. Taylor: "There appears to be definite evidence in the data for a significant deviation from the dipole fit"~\cite{Taylor:1967qv}.  Radiative corrections were also quoted as a point to be treated with particular attention. 

\begin{figure}[h!]
\begin{center}
\includegraphics[width=\columnwidth]{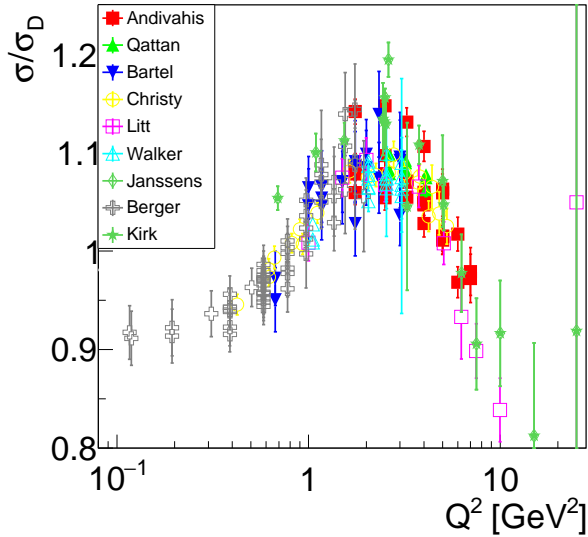}
\caption{Color online. Cross section normalized to the dipole cross section, $\sigma/\sigma_D$,  as a function of $Q^2$ for different experiments:  (Andivahis: \cite{Andivahis:1994rq}, green solid squares), (Qattan: \cite{Qattan:2004ht}, green solid  triangles), (Bartel:  \cite{Bartel:1973rf}, blue triangles down), (Christy: \cite{Christy:2004rc}, yellow open circles), (Litt: \cite{Litt197040}, pink open squares), (Walker: \cite{Walker:1993vj}, cyan open triangles),  (Janssen: \cite{PhysRev.142.922}, green open lozenges) (Berger: \cite{Berger197187}, grey open crosses), (Kirk: \cite{Kirk:1972xm}, green solid stars) compared to the calculation of Ref.~\cite{Bijker:2004yu}, chosen as an example (black solid line).
 }
\label{Fig:dipole}
\end{center}
\end{figure}

The dipole normalized cross section
\be 
\frac{\sigma}{\sigma_D}=\frac{\sigma_{\rm red}^{\rm exp}}{ G_D^2(\epsilon/\mu^2 +\tau)}\,, 
\nonumber\ee
being $\sigma_{\rm red}^{\rm exp}$ the measured reduced cross section, is reported in Fig.~\ref{Fig:dipole} as a function of $Q^2$, regardless of the value of $\epsilon$. The $Q^2$ coordinates for the data  from a Rosenbluth separation for different $\epsilon$  are seen as vertically quasi-aligned symbols. Note that if these points form a cluster with overlapping error bars, it means that they are compatible with the relation $G_E\simeq G_M/\mu\simeq G_D$. If points are not overlapping, then FFs do not follow a dipole behavior. Concerning the data of Ref.~\cite{Andivahis:1994rq}, let us note that the dispersion at fixed $Q^2$ is not larger than the systematics from different sets. 

In general, and particularly at large $Q^2$, one can see that the dipole fit is not a good representation of the data. The deviation at large $Q^2$ reaches 20-30\% on the cross section and has to be attributed mainly to the magnetic term. This is very puzzling, as it is expected that the magnetic FF follows quark counting rules that are compatible with the $Q^2$ dipole dependence.  

\section{Conclusions}

We have proposed a reanalysis of the Rosenbluth data in terms of the squared FF ratio $R^2$ instead that of the extraction of the individual FFs, similarly to what has been done in the time-like region. In such a region, this procedure is more convenient because of the scarce statistics, in the present case it allows to consider $R^2$ as a parameter, directly extracted, avoiding the correlations between $G_E^2$, that is affected by a large error bars, and $G_M^2$ that here includes the eventual systematics and global normalization problems.
We have interpreted our results as follows. In general, the discrepancy between unpolarized and polarized experiments is not evident for the older experiments. Most of them show a decrease of the ratio, already noted in the literature. Up to 3-4 GeV$^2$ the eventual difference may be resolved by a proper calculation of radiative corrections. 

The  claim of the presence of two-photon contributions and the extraction of the FFs  as real quantities, function of one variable, $Q^2$, correcting the unpolarized  the cross section by the assumed effect and re-extracting FFs, is in principle erroneous, as it  integrates the conceptual and operative contradiction of merging the Born approximation and the two-photon effects. In presence of two-photon effects one can not extract nucleon FFs from the unpolarized cross section only.  

In conclusion, in all these analyses the FF extraction is based on the dominance of the one-photon exchange mechanism. Advocating a large contribution of the $1\gamma-2\gamma$ interference, would invalidate the definition of FF itself, as real function of the single variable $Q^2$.
For a specific set of data \cite{Andivahis:1994rq}, we have shown, that, considering two spectrometer settings as independent measurements, brings these data into agreement with the data from the JLab GEp collaboration. Omitting the low energy data also brings the data into agreement, but at the price of increasing the error. The question arises then on the normalization procedure adopted in that paper. The ansatz of constant normalization fixed on one low-$\epsilon$ point on the value that aligns the points at a fixed $Q^2$ is very critical for increasing the Rosenbluth slope. The question remains open for the recent and precise JLab data from Ref.~\cite{Qattan:2004ht}, that show three aligned values for the ratio, increasing with $Q^2$, as well as the data of Refs.~\cite{Walker:1993vj,Litt197040}, showing similar increase although with large errors.

%
%
%
%
\clearpage
\begin{widetext} 
\revappendix
\section{Results for $R^2$ and $G_M^2$}
\label{app:fit}
The obtained values for the parameters $R^2$ and $G_M^2$ in comparison with the corresponding values published in the original analyses~\cite{Andivahis:1994rq,PhysRev.142.922,Bartel:1973rf,Berger197187,Walker:1993vj,Litt197040,Christy:2004rc,Qattan:2004ht} are reported in Tables.~\ref{tab:conf1} and~\ref{tab:conf2}
\begin{table}[h!]
   	\begin{center}\vspace{-0mm}
	\begin{tabular}{c|c|c|c|c|c}
\multirow{3}{*}{Data set} & \multirow{3}{*}{\begin{minipage}{13mm}\centering $q^2$\\(GeV$^2$)\end{minipage}}  & \multicolumn{2}{|c|}{This work} & \multicolumn{2}{|c}{Original} \\
\cline{3-6}
&&
\multirow{2}{*}{$\mu^2\big(R^2\!\pm\! \delta R^2\big)$}
& \multirow{2}{*}{$\displaystyle\frac{G_M^2 \!\pm\! \delta G_M^2}{\mu^2 G_D^2}$} & \multirow{2}{*}{$\mu^2\big(R^2\!\pm\! \delta R^2\big)$}
& \multirow{2}{*}{$\displaystyle\frac{G_M^2 \!\pm\! \delta G_M^2}{\mu^2 G_D^2}$} \\
&  & & & &\\
\hline
 \multirow{7}{*}{\small Ref.~\cite{Andivahis:1994rq}}
 & 1.75 &  0.648$\pm$ 0.089 &  1.140$\pm$ 0.014 &  0.828$\pm$ 0.109 &  1.102$\pm$ 0.021 \\
 & 2.50 &  0.414$\pm$ 0.108 &  1.148$\pm$ 0.014 &  0.679$\pm$ 0.115 &  1.111$\pm$ 0.015 \\
 & 3.25 &  0.260$\pm$ 0.183 &  1.142$\pm$ 0.018 &  0.716$\pm$ 0.200 &  1.092$\pm$ 0.019 \\
 & 4.00 &  0.264$\pm$ 0.211 &  1.111$\pm$ 0.018 &  0.794$\pm$ 0.233 &  1.063$\pm$ 0.019 \\
 & 5.00 &  0.131$\pm$ 0.295 &  1.074$\pm$ 0.018 &  0.867$\pm$ 0.330 &  1.024$\pm$ 0.018 \\
 & 6.00 &  0.073$\pm$ 0.429 &  1.028$\pm$ 0.021 &  0.931$\pm$ 0.426 &  0.974$\pm$ 0.024 \\
 & 7.00 &  1.171$\pm$ 0.841 &  0.971$\pm$ 0.030 &  2.280$\pm$ 0.788 &  0.920$\pm$ 0.031 \\
\hline\hline
 \multirow{21}{*}{\small Ref.~\cite{PhysRev.142.922}}
 & 0.16 &  1.224$\pm$ 0.126 &  0.858$\pm$ 0.050 &  1.223$\pm$ 0.098 &  0.858$\pm$ 0.050 \\
 & 0.18 &  0.975$\pm$ 0.076 &  0.925$\pm$ 0.031 &  1.013$\pm$ 0.060 &  0.918$\pm$ 0.030 \\
 & 0.19 &  0.937$\pm$ 0.129 &  1.007$\pm$ 0.065 &  0.939$\pm$ 0.104 &  1.007$\pm$ 0.068 \\
 & 0.23 &  1.176$\pm$ 0.133 &  0.886$\pm$ 0.048 &  1.172$\pm$ 0.099 &  0.886$\pm$ 0.047 \\
 & 0.27 &  1.130$\pm$ 0.125 &  0.881$\pm$ 0.035 &  1.131$\pm$ 0.102 &  0.881$\pm$ 0.036 \\
 & 0.29 &  1.141$\pm$ 0.145 &  0.884$\pm$ 0.040 &  1.140$\pm$ 0.113 &  0.884$\pm$ 0.041 \\
 & 0.31 &  0.944$\pm$ 0.111 &  0.930$\pm$ 0.036 &  0.945$\pm$ 0.091 &  0.930$\pm$ 0.036 \\
 & 0.35 &  0.933$\pm$ 0.162 &  0.950$\pm$ 0.049 &  0.933$\pm$ 0.128 &  0.950$\pm$ 0.048 \\
 & 0.39 &  1.129$\pm$ 0.123 &  0.918$\pm$ 0.031 &  1.124$\pm$ 0.098 &  0.920$\pm$ 0.032 \\
 & 0.43 &  1.103$\pm$ 0.181 &  0.947$\pm$ 0.047 &  1.103$\pm$ 0.148 &  0.949$\pm$ 0.045 \\
 & 0.47 &  1.041$\pm$ 0.149 &  0.955$\pm$ 0.039 &  1.046$\pm$ 0.122 &  0.953$\pm$ 0.038 \\
 & 0.51 &  1.148$\pm$ 0.242 &  0.918$\pm$ 0.057 &  1.139$\pm$ 0.281 &  0.921$\pm$ 0.084 \\
 & 0.55 &  0.989$\pm$ 0.174 &  0.978$\pm$ 0.041 &  0.994$\pm$ 0.184 &  0.975$\pm$ 0.049 \\
 & 0.58 &  1.057$\pm$ 0.274 &  0.975$\pm$ 0.060 &  1.048$\pm$ 0.380 &  0.973$\pm$ 0.098 \\
 & 0.62 &  0.918$\pm$ 0.197 &  0.987$\pm$ 0.041 &  0.924$\pm$ 0.168 &  0.988$\pm$ 0.042 \\
 & 0.66 &  0.712$\pm$ 0.287 &  1.067$\pm$ 0.065 &  0.714$\pm$ 0.253 &  1.070$\pm$ 0.062 \\
 & 0.70 &  1.199$\pm$ 0.265 &  0.976$\pm$ 0.041 &  1.201$\pm$ 0.233 &  0.975$\pm$ 0.039 \\
 & 0.74 &  1.077$\pm$ 0.386 &  1.044$\pm$ 0.065 &  1.075$\pm$ 0.337 &  1.044$\pm$ 0.068 \\
 & 0.78 &  0.627$\pm$ 0.265 &  1.082$\pm$ 0.046 &  0.623$\pm$ 0.489 &  1.086$\pm$ 0.092 \\
 & 0.86 &  0.555$\pm$ 0.345 &  1.186$\pm$ 0.050 &  0.549$\pm$ 0.497 &  1.189$\pm$ 0.074 \\
\hline\hline
\multirow{8}{*}{\small Ref.~\cite{Bartel:1973rf}}
 & 0.67 &  1.109$\pm$ 0.162 &  0.923$\pm$ 0.049 &  1.110$\pm$ 0.160 &  0.933$\pm$ 0.060 \\
 & 1.00 &  1.108$\pm$ 0.205 &  1.015$\pm$ 0.052 &  1.090$\pm$ 0.210 &  1.036$\pm$ 0.059 \\
 & 1.17 &  0.989$\pm$ 0.153 &  1.041$\pm$ 0.032 &  0.990$\pm$ 0.150 &  1.057$\pm$ 0.043 \\
 & 1.50 &  0.999$\pm$ 0.270 &  1.048$\pm$ 0.056 &  1.000$\pm$ 0.270 &  1.069$\pm$ 0.064 \\
 & 1.75 &  0.608$\pm$ 0.207 &  1.088$\pm$ 0.033 &  0.600$\pm$ 0.210 &  1.118$\pm$ 0.044 \\
 & 2.00 & -4.405$\pm$ 0.010 &  1.525$\pm$ 0.044 &  0.650$\pm$ 0.210 &  1.110$\pm$ 0.042 \\
 & 2.33 & -1.409$\pm$ 0.199 &  1.200$\pm$ 0.045 &  0.510$\pm$ 0.280 &  1.131$\pm$ 0.047 \\
 & 3.00 &  0.409$\pm$ 0.418 &  1.069$\pm$ 0.050 &  0.400$\pm$ 0.320 &  1.115$\pm$ 0.046 \\
\hline\hline
	\end{tabular}
\end{center}
	\caption{\label{tab:conf1} Comparison between our results and published values from Refs.~\cite{Andivahis:1994rq,PhysRev.142.922,Bartel:1973rf}.}
\end{table}
\begin{table}[h!]
   	\begin{center}\vspace{-0mm}
	\begin{tabular}{c|c|c|c|c|c}
\multirow{2}{*}{Dat set} & \multirow{2}{*}{\begin{minipage}{13mm}\centering $q^2$\\(GeV$^2$)\end{minipage}}  & \multicolumn{2}{|c|}{This work} & \multicolumn{2}{|c}{Original} \\
\cline{3-6}
&&
\multirow{2}{*}{$\mu^2\big(R^2\!\pm\! \delta R^2\big)$}
& \multirow{2}{*}{$\displaystyle\frac{G_M^2 \!\pm\! \delta G_M^2}{\mu^2 G_D^2}$} & \multirow{2}{*}{$\mu^2\big(R^2\!\pm\! \delta R^2\big)$}
& \multirow{2}{*}{$\displaystyle\frac{G_M^2 \!\pm\! \delta G_M^2}{\mu^2 G_D^2}$} \\
&  & & & &\\
\hline
 \multirow{10}{*}{\small Ref.~\cite{Berger197187}}
 & 0.12 &  2.603$\pm$13.244 &  0.406$\pm$ 1.855 & - & - \\
 & 0.19 &  1.371$\pm$ 3.302 &  0.734$\pm$ 1.296 & - & - \\
 & 0.39 &  0.915$\pm$ 0.083 &  0.973$\pm$ 0.037 &  0.943$\pm$ 0.089 &  0.960$\pm$ 0.039 \\
 & 0.58 &  0.977$\pm$ 0.054 &  0.973$\pm$ 0.017 &  0.976$\pm$ 0.053 &  0.973$\pm$ 0.017 \\
 & 0.78 &  0.902$\pm$ 0.097 &  1.014$\pm$ 0.025 &  0.910$\pm$ 0.076 &  1.013$\pm$ 0.022 \\
 & 0.97 &  1.065$\pm$ 0.152 &  1.015$\pm$ 0.033 &  0.949$\pm$ 0.103 &  1.033$\pm$ 0.027 \\
 & 1.17 &  1.099$\pm$ 0.318 &  1.057$\pm$ 0.048 &  0.762$\pm$ 0.122 &  1.102$\pm$ 0.032 \\
 & 1.36 &  0.787$\pm$ 0.320 &  1.085$\pm$ 0.046 &  0.832$\pm$ 0.146 &  1.081$\pm$ 0.034 \\
 & 1.56 &  1.444$\pm$ 0.683 &  1.073$\pm$ 0.057 &  0.686$\pm$ 0.151 &  1.130$\pm$ 0.035 \\
 & 1.75 &  0.661$\pm$ 0.776 &  1.142$\pm$ 0.078 &  0.590$\pm$ 0.194 &  1.148$\pm$ 0.044 \\
 \hline\hline
		\multirow{4}{*}{\small Ref.~\cite{Walker:1993vj}}
 & 1.00 &  0.991$\pm$ 0.486 &  1.023$\pm$ 0.136 &  0.969$\pm$ 0.172 &  1.034$\pm$ 0.066 \\
 & 2.00 &  1.311$\pm$ 0.467 &  1.024$\pm$ 0.073 &  1.338$\pm$ 0.224 &  1.028$\pm$ 0.044 \\
 & 2.50 &  1.336$\pm$ 0.583 &  1.030$\pm$ 0.074 &  1.143$\pm$ 0.265 &  1.061$\pm$ 0.044 \\
 & 3.01 &  1.563$\pm$ 0.777 &  1.008$\pm$ 0.080 &  1.480$\pm$ 0.409 &  1.024$\pm$ 0.050 \\
\hline\hline
\multirow{5}{*}{\small Ref.~\cite{Litt197040}}
 & 1.00 & -0.587$\pm$ 0.693 &  1.926$\pm$ 0.750 &  0.941$\pm$ 0.097 & - \\
 & 1.50 &  1.581$\pm$ 1.218 &  0.953$\pm$ 0.221 &  0.672$\pm$ 0.131 & - \\
 & 2.00 &  1.665$\pm$ 0.997 &  0.969$\pm$ 0.142 &  1.124$\pm$ 0.360 & - \\
 & 2.50 &  1.170$\pm$ 0.419 &  1.039$\pm$ 0.057 &  1.346$\pm$ 0.441 & - \\
 & 3.75 &  2.308$\pm$ 0.862 &  0.957$\pm$ 0.073 &  1.988$\pm$ 0.874 & - \\
\hline\hline
	 \multirow{7}{*}{\small Ref.~\cite{Christy:2004rc}}
 & 0.65 &  1.144$\pm$ 0.118 &  0.937$\pm$ 0.039 &  1.143$\pm$ 0.182 &  1.071$\pm$ 0.108 \\
 & 0.90 &  0.830$\pm$ 0.011 &  1.065$\pm$ 0.004 &  0.861$\pm$ 0.124 &  0.910$\pm$ 0.101 \\
 & 2.20 &  0.766$\pm$ 0.107 &  1.104$\pm$ 0.016 &  0.771$\pm$ 0.220 &  0.852$\pm$ 0.223 \\
 & 2.75 &  0.709$\pm$ 0.153 &  1.113$\pm$ 0.017 &  0.707$\pm$ 0.183 &  0.789$\pm$ 0.202 \\
 & 3.75 &  0.695$\pm$ 0.643 &  1.090$\pm$ 0.051 &  0.701$\pm$ 0.368 &  0.762$\pm$ 0.405 \\
 & 4.25 &  1.526$\pm$ 0.961 &  1.026$\pm$ 0.058 &  1.538$\pm$ 0.404 &  1.575$\pm$ 0.394 \\
 & 5.25 &  1.382$\pm$ 0.508 &  1.014$\pm$ 0.026 &  1.383$\pm$ 1.298 &  1.399$\pm$ 1.209 \\
\hline\hline
		 \multirow{3}{*}{\small Ref.~\cite{Qattan:2004ht}}
 & 2.64 &  0.817$\pm$ 0.058 &  1.104$\pm$ 0.006 &  0.814$\pm$ 0.069 &  1.109$\pm$ 0.032 \\
 & 3.20 &  0.947$\pm$ 0.087 &  1.093$\pm$ 0.007 &  0.924$\pm$ 0.098 &  1.098$\pm$ 0.031 \\
 & 4.10 &  1.266$\pm$ 0.161 &  1.058$\pm$ 0.009 &  1.203$\pm$ 0.169 &  1.063$\pm$ 0.031 \\
\hline\hline
	\end{tabular}\end{center}
	\caption{\label{tab:conf2} Comparison between our results  and published values from Refs.~\cite{Bartel:1973rf,Berger197187,Walker:1993vj,Litt197040,Christy:2004rc,Qattan:2004ht}}
\end{table}
\clearpage
\section{Fit of the data of Ref.~\cite{Andivahis:1994rq}}
\label{app:andi}
Table~\ref{tab:andi} reports the original data from Ref.~\cite{Andivahis:1994rq}. Fits to these data, for the seven 
$Q^2$ values, are shown in Fig.~\ref{fig:andi}.
%
%
\begin{table}[h!]
\begin{center}
\begin{tabular}{c|c|c||c|c|c}
$Q^2$  & \multirow{2}{*}{$\epsilon$} & \multirow{2}{*}{$\sigma^{\rm red}$} &
	$Q^2$  & \multirow{2}{*}{$\epsilon$} & \multirow{2}{*}{$\sigma^{\rm red}$}\\
(GeV$^2$) & & & (GeV$^2$) & &\\
	\hline
1.75 & 0.250 & 0.032703 $\pm$ 0.000365 & 4.00 & 0.190 & 0.005172 $\pm$ 0.000071 \\
1.75 & 0.250 & 0.031105 $\pm$ 0.000378 & 4.00 & 0.437 & 0.005063 $\pm$ 0.000088 \\
1.75 & 0.704 & 0.034287 $\pm$ 0.000363 & 4.00 & 0.593 & 0.005117 $\pm$ 0.000081 \\
1.75 & 0.950 & 0.035499 $\pm$ 0.000411 & 4.00 & 0.694 & 0.005231 $\pm$ 0.000083 \\
 \cline{1-3}
2.50 & 0.227 & 0.015839 $\pm$ 0.000181 & 4.00 & 0.805 & 0.005133 $\pm$ 0.000069 \\
2.50 & 0.227 & 0.015142 $\pm$ 0.000216 & 4.00 & 0.946 & 0.005303 $\pm$ 0.000069 \\
 \cline{4-6}
2.50 & 0.479 & 0.015540 $\pm$ 0.000208 & 5.00 & 0.171 & 0.002859 $\pm$ 0.000041 \\
2.50 & 0.630 & 0.015766 $\pm$ 0.000209 & 5.00 & 0.389 & 0.002904 $\pm$ 0.000066 \\
2.50 & 0.750 & 0.016112 $\pm$ 0.000174 & 5.00 & 0.538 & 0.002819 $\pm$ 0.000050 \\
2.50 & 0.820 & 0.016239 $\pm$ 0.000187 & 5.00 & 0.704 & 0.002849 $\pm$ 0.000041 \\
2.50 & 0.913 & 0.016282 $\pm$ 0.000194 & 5.00 & 0.919 & 0.002902 $\pm$ 0.000043 \\
 \cline{1-3}
 \cline{4-6}
3.25 & 0.206 & 0.008660 $\pm$ 0.000112 & 6.00 & 0.156 & 0.001715 $\pm$ 0.000029 \\
3.25 & 0.426 & 0.008438 $\pm$ 0.000132 & 6.00 & 0.886 & 0.001721 $\pm$ 0.000028 \\
 \cline{4-6}
3.25 & 0.609 & 0.008646 $\pm$ 0.000113 & 7.00 & 0.847 & 0.001152 $\pm$ 0.000029 \\
3.25 & 0.719 & 0.008759 $\pm$ 0.000114 & 7.00 & 0.143 & 0.001094 $\pm$ 0.000028 \\
 \cline{4-6}
3.25 & 0.865 & 0.008793 $\pm$ 0.000098 & 8.83 & 0.125 & 0.000530 $\pm$ 0.000022 \\
 \cline{1-3}
 \cline{4-6}
\end{tabular}	
\caption{\label{tab:andi}Reduced cross section from Ref.~\cite{Andivahis:1994rq}.}
\end{center}
\end{table}
%
%
\begin{figure}[h!]
\begin{center}
\includegraphics[width=\textwidth]{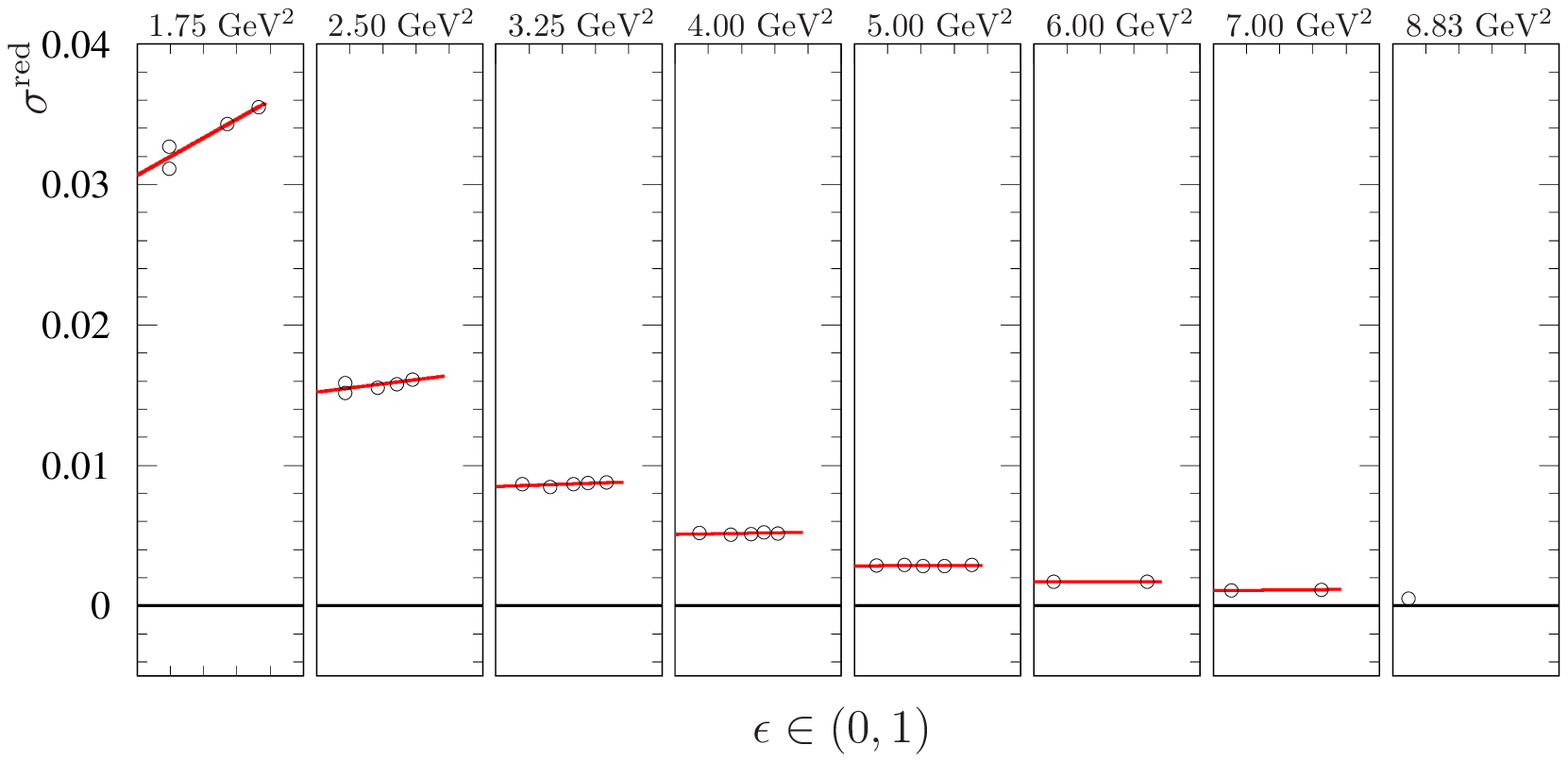}
\vspace{-10mm}\\
\caption{The graphs represent data (open circles) form Ref.~\cite{Andivahis:1994rq} and fits (red lines) 
of the reduced cross section as a function of $\epsilon$, at the $Q^2$ values reported on the top of each graph.}
\label{fig:andi}
\end{center}\end{figure}
\clearpage
\section{Fit of the data of Ref.~\cite{PhysRev.142.922}}
\label{app:jans}
Table~\ref{tab:jans} reports the original data from Ref.~\cite{PhysRev.142.922}. Fits to these data, for the 20 
$Q^2$ values, are shown in Fig.~\ref{fig:jans}.
%
%
\begin{table}[h]
\scalebox{1.1}{\scriptsize
\begin{tabular}{c|c|c||c|c|c||c|c|c}
	$Q^2$  & \multirow{2}{*}{$\epsilon$} & \multirow{2}{*}{$\sigma^{\rm red}$} &
	$Q^2$  & \multirow{2}{*}{$\epsilon$} & \multirow{2}{*}{$\sigma^{\rm red}$} &
	$Q^2$  & \multirow{2}{*}{$\epsilon$} & \multirow{2}{*}{$\sigma^{\rm red}$} \\	
	(GeV$^2$) &  & &
	(GeV$^2$) &  & &
	(GeV$^2$) &  &\\
	\hline
0.16 & 0.736 & 0.4882 $\pm$ 0.0192 & 0.31 & 0.044 & 0.1640 $\pm$ 0.0082 & 0.62 & 0.541 & 0.1514 $\pm$ 0.0076 \\
 \cline{4-6}
0.16 & 0.449 & 0.3433 $\pm$ 0.0136 & 0.35 & 0.577 & 0.2487 $\pm$ 0.0127 & 0.62 & 0.492 & 0.1477 $\pm$ 0.0075 \\
0.16 & 0.076 & 0.1704 $\pm$ 0.0067 & 0.35 & 0.436 & 0.2276 $\pm$ 0.0091 & 0.62 & 0.419 & 0.1332 $\pm$ 0.0064 \\
 \cline{1-3}
0.18 & 0.735 & 0.4356 $\pm$ 0.0178 & 0.35 & 0.072 & 0.1609 $\pm$ 0.0065 & 0.62 & 0.376 & 0.1408 $\pm$ 0.0071 \\
 \cline{4-6}
0.18 & 0.588 & 0.3448 $\pm$ 0.0148 & 0.39 & 0.709 & 0.2747 $\pm$ 0.0134 & 0.62 & 0.068 & 0.1139 $\pm$ 0.0057 \\
0.18 & 0.447 & 0.3180 $\pm$ 0.0127 & 0.39 & 0.575 & 0.2387 $\pm$ 0.0096 & 0.62 & 0.041 & 0.1135 $\pm$ 0.0057 \\
 \cline{7-9}
0.18 & 0.322 & 0.2676 $\pm$ 0.0132 & 0.39 & 0.433 & 0.2117 $\pm$ 0.0087 & 0.66 & 0.506 & 0.1430 $\pm$ 0.0072 \\
0.18 & 0.137 & 0.1959 $\pm$ 0.0098 & 0.39 & 0.310 & 0.1955 $\pm$ 0.0097 & 0.66 & 0.417 & 0.1317 $\pm$ 0.0064 \\
0.18 & 0.075 & 0.1808 $\pm$ 0.0072 & 0.39 & 0.131 & 0.1594 $\pm$ 0.0080 & 0.66 & 0.067 & 0.1164 $\pm$ 0.0058 \\
 \cline{7-9}
0.18 & 0.045 & 0.1642 $\pm$ 0.0066 & 0.39 & 0.072 & 0.1505 $\pm$ 0.0060 & 0.68 & 0.416 & 0.1325 $\pm$ 0.0055 \\
 \cline{1-3}
 \cline{7-9}
0.19 & 0.776 & 0.4308 $\pm$ 0.0213 & 0.39 & 0.043 & 0.1480 $\pm$ 0.0074 & 0.70 & 0.470 & 0.1375 $\pm$ 0.0068 \\
 \cline{4-6}
0.19 & 0.446 & 0.3388 $\pm$ 0.0169 & 0.43 & 0.707 & 0.2536 $\pm$ 0.0127 & 0.70 & 0.415 & 0.1288 $\pm$ 0.0065 \\
0.19 & 0.075 & 0.1896 $\pm$ 0.0094 & 0.43 & 0.431 & 0.1981 $\pm$ 0.0097 & 0.70 & 0.372 & 0.1200 $\pm$ 0.0058 \\
 \cline{1-3}
0.23 & 0.727 & 0.4055 $\pm$ 0.0202 & 0.43 & 0.071 & 0.1482 $\pm$ 0.0059 & 0.70 & 0.067 & 0.1021 $\pm$ 0.0051 \\
 \cline{4-6}
0.23 & 0.585 & 0.3366 $\pm$ 0.0137 & 0.47 & 0.675 & 0.2248 $\pm$ 0.0115 & 0.70 & 0.040 & 0.1012 $\pm$ 0.0051 \\
 \cline{7-9}
0.23 & 0.443 & 0.2888 $\pm$ 0.0146 & 0.47 & 0.639 & 0.2158 $\pm$ 0.0109 & 0.74 & 0.434 & 0.1229 $\pm$ 0.0063 \\
0.23 & 0.074 & 0.1727 $\pm$ 0.0069 & 0.47 & 0.428 & 0.1830 $\pm$ 0.0073 & 0.74 & 0.412 & 0.1312 $\pm$ 0.0080 \\
 \cline{1-3}
0.27 & 0.582 & 0.3072 $\pm$ 0.0125 & 0.47 & 0.070 & 0.1396 $\pm$ 0.0069 & 0.74 & 0.066 & 0.1025 $\pm$ 0.0051 \\
 \cline{7-9}
0.27 & 0.441 & 0.2707 $\pm$ 0.0109 & 0.47 & 0.042 & 0.1381 $\pm$ 0.0069 & 0.78 & 0.410 & 0.1197 $\pm$ 0.0060 \\
 \cline{4-6}
0.27 & 0.280 & 0.2060 $\pm$ 0.0103 & 0.49 & 0.427 & 0.1722 $\pm$ 0.0068 & 0.78 & 0.397 & 0.1213 $\pm$ 0.0058 \\
 \cline{4-6}
0.27 & 0.074 & 0.1637 $\pm$ 0.0065 & 0.51 & 0.642 & 0.2079 $\pm$ 0.0103 & 0.78 & 0.368 & 0.1015 $\pm$ 0.0049 \\
0.27 & 0.044 & 0.1635 $\pm$ 0.0082 & 0.51 & 0.426 & 0.1647 $\pm$ 0.0068 & 0.78 & 0.337 & 0.0996 $\pm$ 0.0049 \\
 \cline{1-3}
0.29 & 0.581 & 0.2924 $\pm$ 0.0120 & 0.51 & 0.070 & 0.1305 $\pm$ 0.0065 & 0.78 & 0.066 & 0.1098 $\pm$ 0.0054 \\
 \cline{4-6}
0.29 & 0.440 & 0.2570 $\pm$ 0.0105 & 0.55 & 0.609 & 0.1955 $\pm$ 0.0098 & 0.78 & 0.039 & 0.0927 $\pm$ 0.0046 \\
 \cline{7-9}
0.29 & 0.316 & 0.2234 $\pm$ 0.0113 & 0.55 & 0.567 & 0.1774 $\pm$ 0.0092 & 0.86 & 0.406 & 0.1130 $\pm$ 0.0056 \\
0.29 & 0.133 & 0.1733 $\pm$ 0.0087 & 0.55 & 0.424 & 0.1542 $\pm$ 0.0064 & 0.86 & 0.324 & 0.0945 $\pm$ 0.0056 \\
0.29 & 0.073 & 0.1653 $\pm$ 0.0066 & 0.55 & 0.069 & 0.1258 $\pm$ 0.0063 & 0.86 & 0.065 & 0.0995 $\pm$ 0.0050 \\
 \cline{1-3}
0.31 & 0.626 & 0.2746 $\pm$ 0.0136 & 0.55 & 0.041 & 0.1301 $\pm$ 0.0065 & 0.86 & 0.038 & 0.0951 $\pm$ 0.0047 \\
 \cline{4-6}
 \cline{7-9}
0.31 & 0.580 & 0.2674 $\pm$ 0.0132 & 0.58 & 0.575 & 0.1770 $\pm$ 0.0087 & 1.01 & 0.037 & 0.0725 $\pm$ 0.0039 \\
 \cline{7-9}
0.31 & 0.438 & 0.2433 $\pm$ 0.0095 & 0.58 & 0.421 & 0.1444 $\pm$ 0.0074 & 1.09 & 0.037 & 0.0630 $\pm$ 0.0044 \\
 \cline{7-9}
0.31 & 0.073 & 0.1606 $\pm$ 0.0064 & 0.58 & 0.069 & 0.1227 $\pm$ 0.0062 & 1.17 & 0.036 & 0.0551 $\pm$ 0.0044 \\
 \cline{4-6}
 \cline{7-9}
\end{tabular}t}
\caption{\label{tab:jans}Reduced cross section from Ref.~\cite{PhysRev.142.922}.}
\end{table}\vspace{2mm}
\newpage
%
%
\begin{figure}[h!]
\begin{center}
\includegraphics[width=\textwidth]{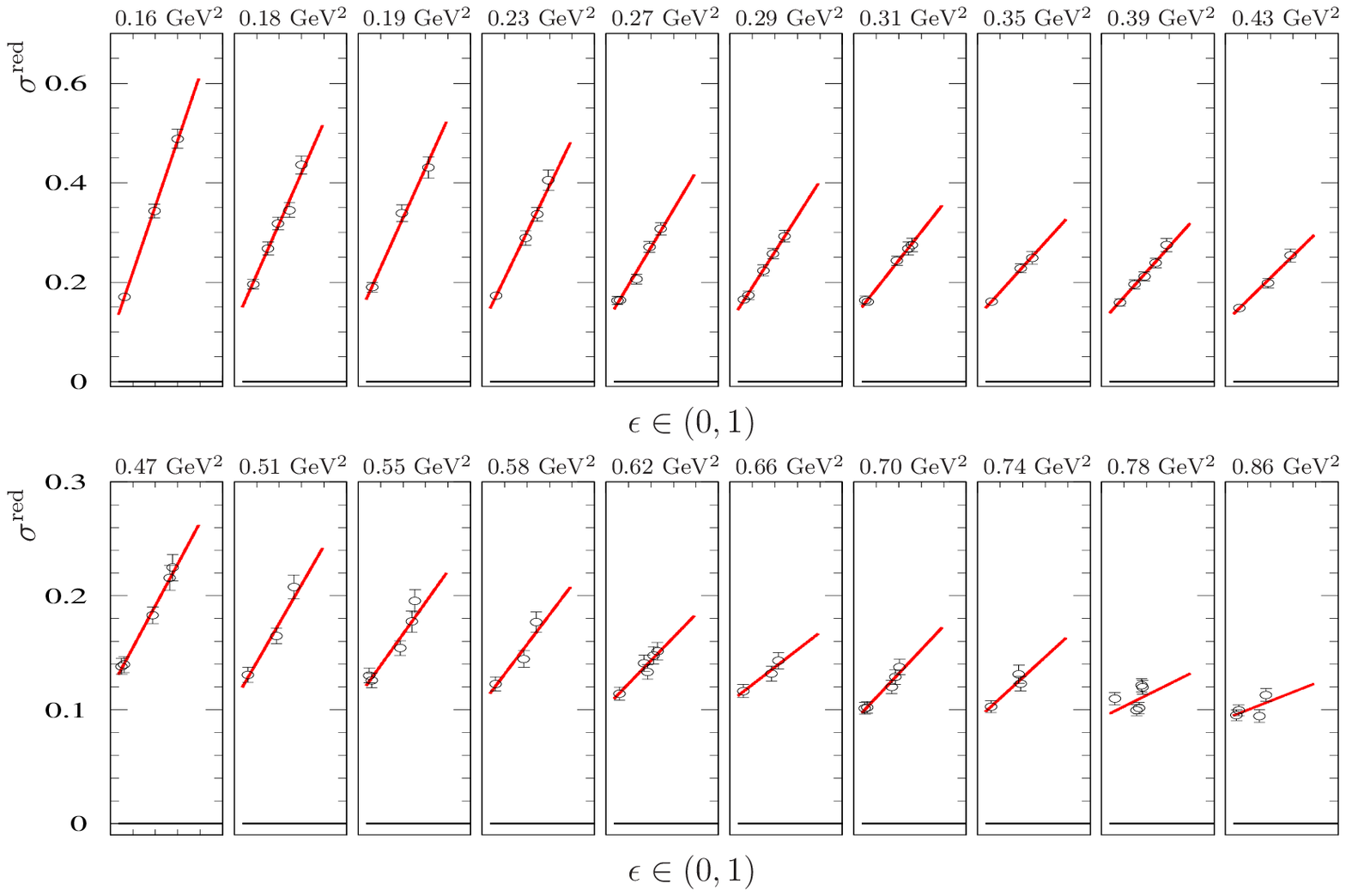}\vspace{-2mm}
\caption{\label{fig:jans}The graphs represent data (open circles) from Ref.~\cite{PhysRev.142.922} and fits (red lines) 
of the reduced cross section as a function of $\epsilon$, at the $Q^2$ values reported on the top of each graph.}
\end{center}
\vspace{-15mm}
\end{figure}
\clearpage
\section{Fit of the data of Ref.~\cite{Bartel:1973rf}}
\label{app:bart}
Table~\ref{tab:bart} reports the original data from Ref.~\cite{Bartel:1973rf}. Fits to these data, for the eight 
$Q^2$ values, are shown in Fig.~\ref{fig:bart}.
%
%
\begin{table}[h]
\begin{center}
	\begin{tabular}{c|c|c||c|c|c}
	$Q^2$ & \multirow{2}{*}{$\epsilon$} & \multirow{2}{*}{$\sigma^{\rm red}$} &
	$Q^2$ & \multirow{2}{*}{$\epsilon$} & \multirow{2}{*}{$\sigma^{\rm red}$}\\
	(GeV$^2$) &&&(GeV$^2$) &&\\
	\hline
	0.67 & 0.974 & 0.16589 $\pm$ 0.00315 & 1.75 & 0.965 & 0.03371 $\pm$ 0.00081 \\
0.67 & 0.326 & 0.11937 $\pm$ 0.00322 & 1.75 & 0.250 & 0.03048 $\pm$ 0.00098 \\
 \cline{1-3}
1.00 & 0.972 & 0.09933 $\pm$ 0.00209 & 1.75 & 0.278 & 0.03052 $\pm$ 0.00076 \\
 \cline{4-6}
1.00 & 0.309 & 0.07717 $\pm$ 0.00224 & 2.00 & 0.948 & 0.00181 $\pm$ 0.00004 \\
 \cline{1-3}
1.17 & 0.969 & 0.07534 $\pm$ 0.00143 & 2.00 & 0.268 & 0.02334 $\pm$ 0.00065 \\
 \cline{4-6}
1.17 & 0.273 & 0.06127 $\pm$ 0.00165 & 2.33 & 0.952 & 0.01364 $\pm$ 0.00035 \\
1.17 & 0.301 & 0.06081 $\pm$ 0.00152 & 2.33 & 0.257 & 0.01714 $\pm$ 0.00050 \\
 \cline{1-3}
 \cline{4-6}
1.50 & 0.970 & 0.04791 $\pm$ 0.00096 & 3.00 & 0.918 & 0.01006 $\pm$ 0.00021 \\
1.50 & 0.287 & 0.04030 $\pm$ 0.00137 & 3.00 & 0.237 & 0.00966 $\pm$ 0.00033 \\
 \cline{1-3}
 \cline{4-6}
\end{tabular}	
\caption{\label{tab:bart}Reduced cross section from Ref.~\cite{Bartel:1973rf}.}
\end{center}
\end{table}\vspace{4mm}
%
%
\begin{figure}[h!]
\begin{center}
\includegraphics[width=\textwidth]{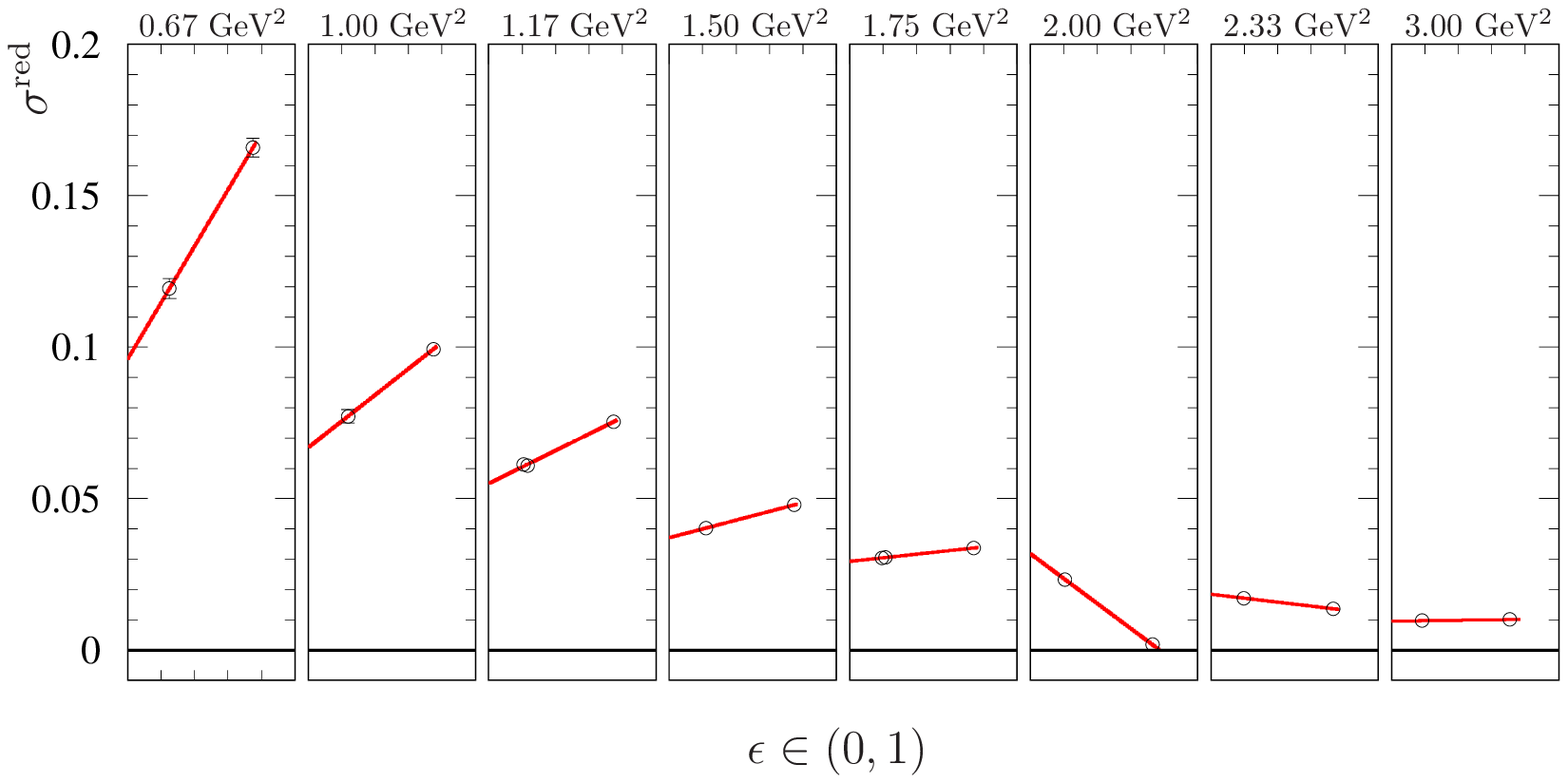}	\vspace{-2mm}
\caption{\label{fig:bart} The graphs represent data (open circles) from Ref.~\cite{Bartel:1973rf} and fits (red lines) 
of the reduced cross section as a function of $\epsilon$, at the $Q^2$ values reported on the top of each graph.}
\end{center}
\end{figure}
\clearpage
\section{Fit of the data of Ref.~\cite{Berger197187}}
\label{app:berg}
Table~\ref{tab:berg} reports the original data from Ref.~\cite{Berger197187}. Fits to these data, for the eight 
$Q^2$ values, are shown in Fig.~\ref{fig:berg}.

%
%
%
\begin{table}[h]\begin{center}\vspace{-0mm}
\scalebox{1}{\scriptsize
\begin{tabular}{c|c|c||c|c|c||c|c|c}
	$Q^2$  & \multirow{2}{*}{$\epsilon$} & \multirow{2}{*}{$\sigma^{\rm red}$} &
	$Q^2$  & \multirow{2}{*}{$\epsilon$} & \multirow{2}{*}{$\sigma^{\rm red}$} &
	$Q^2$ & \multirow{2}{*}{$\epsilon$} & \multirow{2}{*}{$\sigma^{\rm red}$}\\
(GeV$^2$) &&&(GeV$^2$) &&&(GeV$^2$) &&\\
	\hline
	0.08 & 0.907 & 0.6535 $\pm$ 0.0197 & 0.58 & 0.810 & 0.1858 $\pm$ 0.0037 & 0.97 & 0.499 & 0.0862 $\pm$ 0.0020 \\
 \cline{1-3}
0.12 & 0.906 & 0.5782 $\pm$ 0.0171 & 0.58 & 0.720 & 0.1740 $\pm$ 0.0035 & 0.97 & 0.281 & 0.0801 $\pm$ 0.0020 \\
0.12 & 0.828 & 0.5334 $\pm$ 0.0161 & 0.58 & 0.663 & 0.1684 $\pm$ 0.0034 & 0.97 & 0.161 & 0.0732 $\pm$ 0.0025 \\
 \cline{1-3}
 \cline{7-9}
0.19 & 0.904 & 0.4656 $\pm$ 0.0144 & 0.58 & 0.644 & 0.1713 $\pm$ 0.0035 & 1.17 & 0.736 & 0.0729 $\pm$ 0.0029 \\
0.19 & 0.825 & 0.4354 $\pm$ 0.0131 & 0.58 & 0.522 & 0.1596 $\pm$ 0.0032 & 1.17 & 0.489 & 0.0679 $\pm$ 0.0015 \\
 \cline{1-3}
0.31 & 0.901 & 0.3471 $\pm$ 0.0086 & 0.58 & 0.516 & 0.1573 $\pm$ 0.0032 & 1.17 & 0.272 & 0.0616 $\pm$ 0.0016 \\
 \cline{1-3}
0.39 & 0.900 & 0.2884 $\pm$ 0.0058 & 0.58 & 0.300 & 0.1401 $\pm$ 0.0031 & 1.17 & 0.155 & 0.0611 $\pm$ 0.0031 \\
 \cline{7-9}
0.39 & 0.848 & 0.2727 $\pm$ 0.0053 & 0.58 & 0.298 & 0.1402 $\pm$ 0.0028 & 1.36 & 0.623 & 0.0518 $\pm$ 0.0014 \\
0.39 & 0.843 & 0.2788 $\pm$ 0.0055 & 0.58 & 0.173 & 0.1274 $\pm$ 0.0029 & 1.36 & 0.473 & 0.0522 $\pm$ 0.0019 \\
0.39 & 0.818 & 0.2758 $\pm$ 0.0056 & 0.58 & 0.170 & 0.1302 $\pm$ 0.0027 & 1.36 & 0.263 & 0.0485 $\pm$ 0.0016 \\
 \cline{4-6}
0.39 & 0.674 & 0.2468 $\pm$ 0.0050 & 0.78 & 0.826 & 0.1293 $\pm$ 0.0027 & 1.36 & 0.147 & 0.0461 $\pm$ 0.0018 \\
 \cline{7-9}
0.39 & 0.529 & 0.2274 $\pm$ 0.0037 & 0.78 & 0.803 & 0.1289 $\pm$ 0.0038 & 1.56 & 0.469 & 0.0431 $\pm$ 0.0016 \\
0.39 & 0.308 & 0.1947 $\pm$ 0.0038 & 0.78 & 0.710 & 0.1259 $\pm$ 0.0025 & 1.56 & 0.257 & 0.0380 $\pm$ 0.0016 \\
 \cline{1-3}
0.51 & 0.897 & 0.2256 $\pm$ 0.0046 & 0.78 & 0.510 & 0.1114 $\pm$ 0.0027 & 1.56 & 0.145 & 0.0384 $\pm$ 0.0013 \\
 \cline{1-3}
 \cline{7-9}
0.58 & 0.895 & 0.1899 $\pm$ 0.0038 & 0.78 & 0.290 & 0.1054 $\pm$ 0.0023 & 1.75 & 0.454 & 0.0328 $\pm$ 0.0010 \\
0.58 & 0.855 & 0.1923 $\pm$ 0.0038 & 0.78 & 0.167 & 0.0984 $\pm$ 0.0021 & 1.75 & 0.248 & 0.0328 $\pm$ 0.0013 \\
 \cline{4-6}
0.58 & 0.837 & 0.1866 $\pm$ 0.0036 & 0.97 & 0.796 & 0.0949 $\pm$ 0.0019 & 1.75 & 0.138 & 0.0303 $\pm$ 0.0018 \\
 \cline{7-9}
0.58 & 0.833 & 0.1836 $\pm$ 0.0037 & 0.97 & 0.745 & 0.0968 $\pm$ 0.0021 & 1.95 & 0.243 & 0.0246 $\pm$ 0.0013 \\
 \cline{7-9}
\end{tabular}}
\caption{\label{tab:berg}Reduced cross section from Ref.~\cite{Berger197187}.}
\end{center}\vspace{2mm}\end{table}
%
%
\begin{figure}[h!]
\begin{center}
\includegraphics[width=\textwidth]{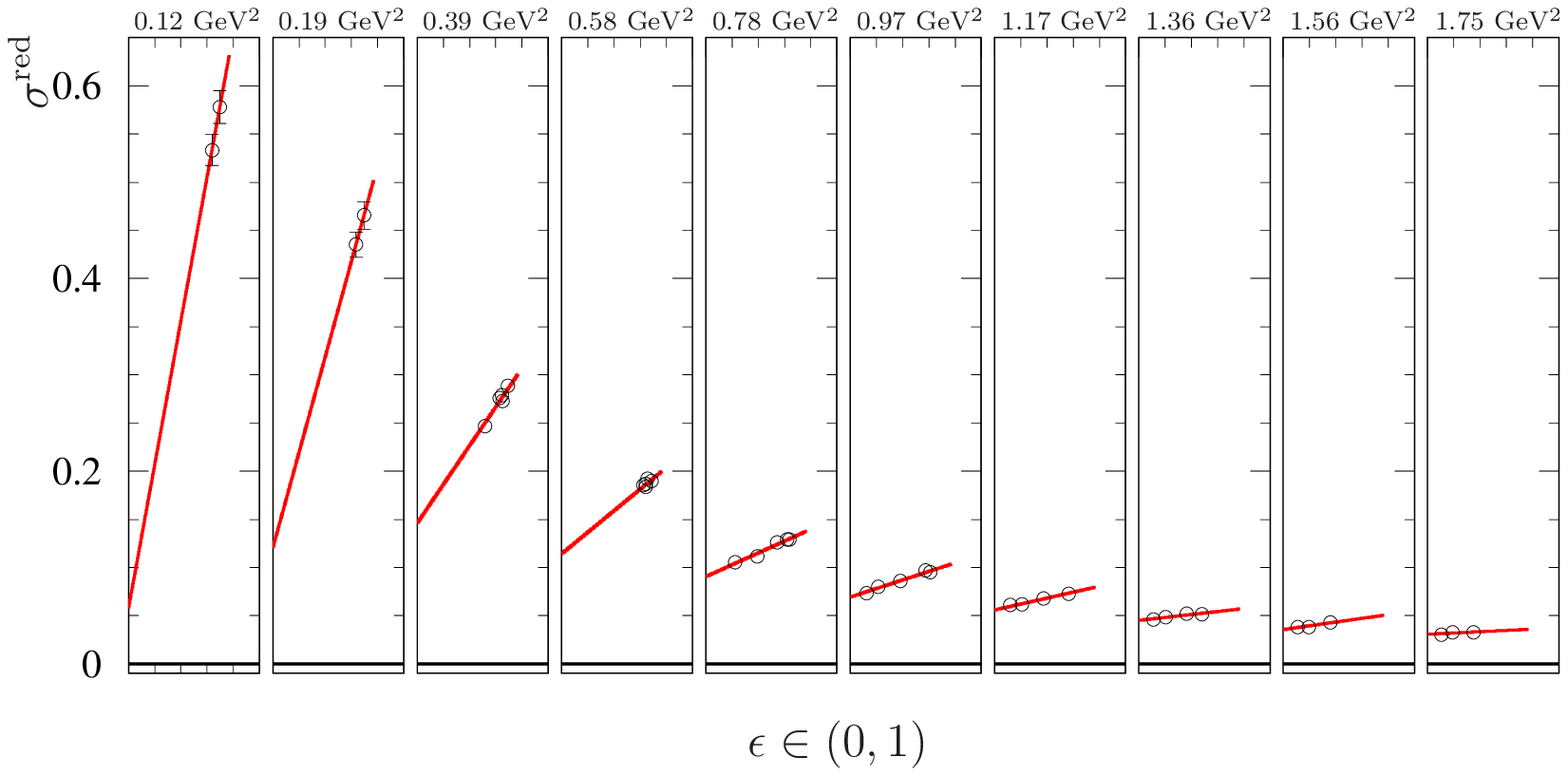}	\vspace{-2mm}
\caption{\label{fig:berg}The graphs represent data (open circles) form Ref.~\cite{Berger197187} and fits (red lines) 
of the reduced cross section as a function of $\epsilon$, at the $Q^2$ values reported on the top of each graph.}
\end{center}
\end{figure}
\clearpage
\section{Fit of the data of Ref.~\cite{Christy:2004rc}}
\label{app:chri}
Table~\ref{tab:chri} reports the original data from Ref.~\cite{Christy:2004rc}. Fits to these data, for the seven 
$Q^2$ values, are shown in Fig.~\ref{fig:chri}.
%
%
\begin{table}[h]
\begin{center}
	\begin{tabular}{c|c|c||c|c|c}
	$Q^2$  & \multirow{2}{*}{$\epsilon$} & \multirow{2}{*}{$\sigma^{\rm red}$} &
	$Q^2$  & \multirow{2}{*}{$\epsilon$} & \multirow{2}{*}{$\sigma^{\rm red}$}\\
	(GeV$^2$) &&&(GeV$^2$)&&\\
	\hline
	0.65 & 0.682 & 0.15452 $\pm$ 0.00099 & 3.75 & 0.403 & 0.00599 $\pm$ 0.00012 \\
0.65 & 0.919 & 0.17305 $\pm$ 0.00108 & 3.75 & 0.658 & 0.00620 $\pm$ 0.00012 \\
0.65 & 0.955 & 0.17646 $\pm$ 0.00115 & 3.75 & 0.826 & 0.00616 $\pm$ 0.00014 \\
 \cline{1-3}
 \cline{4-6}
0.90 & 0.549 & 0.09835 $\pm$ 0.00010 & 4.25 & 0.257 & 0.00421 $\pm$ 0.00015 \\
0.90 & 0.810 & 0.10843 $\pm$ 0.00010 & 4.25 & 0.553 & 0.00448 $\pm$ 0.00021 \\
0.90 & 0.931 & 0.11098 $\pm$ 0.00008 & 4.25 & 0.786 & 0.00456 $\pm$ 0.00013 \\
 \cline{1-3}
 \cline{4-6}
2.20 & 0.488 & 0.02048 $\pm$ 0.00011 & 5.25 & 0.469 & 0.00250 $\pm$ 0.00002 \\
2.20 & 0.783 & 0.02158 $\pm$ 0.00010 & 5.25 & 0.702 & 0.00255 $\pm$ 0.00001 \\
2.20 & 0.924 & 0.02168 $\pm$ 0.00013 & 5.25 & 0.659 & 0.00259 $\pm$ 0.00002 \\
 \cline{1-3}
 \cline{4-6}
2.75 & 0.284 & 0.01243 $\pm$ 0.00012 &\multicolumn{3}{c}{}  \\
2.75 & 0.673 & 0.01295 $\pm$ 0.00011 &\multicolumn{3}{c}{}  \\
2.75 & 0.896 & 0.01329 $\pm$ 0.00012 &\multicolumn{3}{c}{}  \\
 \cline{1-3}
\end{tabular}	
\caption{Reduced cross section from Ref.~\cite{Christy:2004rc}.}
\label{tab:chri}
\end{center}
\vspace{-5mm}\end{table}
%
%
\begin{figure}[h!]
\begin{center}
\includegraphics[width=\textwidth]{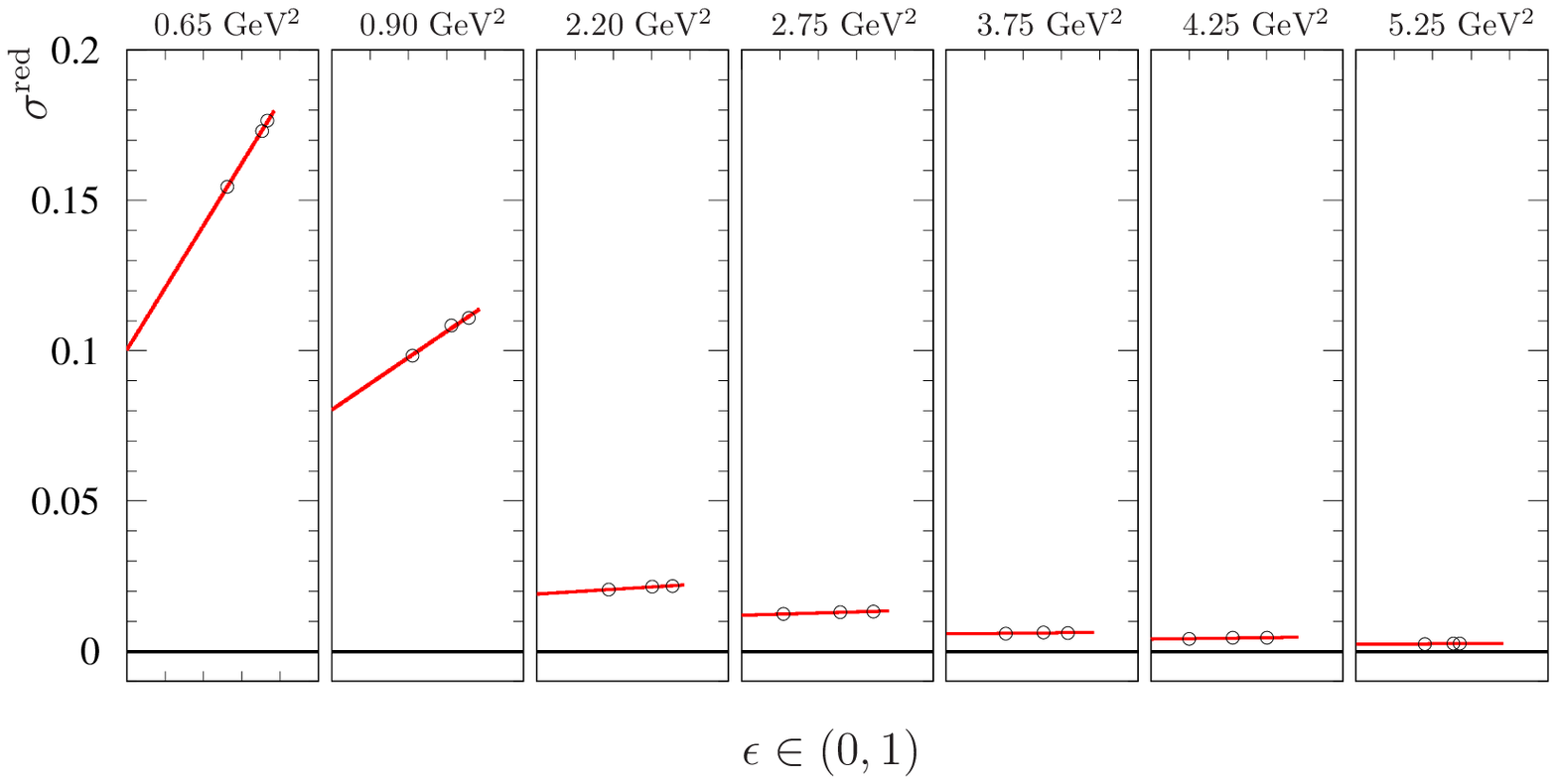}\vspace{-2mm}
\caption{\label{fig:chri}The graphs represent data (open circles) from Ref.~\cite{Christy:2004rc} and fits (red lines) 
of the reduced cross section as a function of $\epsilon$, at the $Q^2$ values reported on the top of each graph.}\end{center}
\end{figure}
\clearpage

\section{Fit of the data of Ref.~\cite{Walker:1993vj}}
\label{app:walk}
Table~\ref{tab:walk} reports the original data from Ref.~\cite{Walker:1993vj}. Fits to these data, for the four 
$Q^2$ values, are shown in Fig.~\ref{fig:walk}.
%
%
\begin{table}[h]
\begin{center}
\begin{tabular}{c|c|c||c|c|c}
	$Q^2$  & \multirow{2}{*}{$\epsilon$} & \multirow{2}{*}{$\sigma^{\rm red}$} &
	$Q^2$  & \multirow{2}{*}{$\epsilon$} & \multirow{2}{*}{$\sigma^{\rm red}$}\\
	(GeV$^2$) &&&(GeV$^2$)&&\\
	\hline
	1.00 & 0.692 & 0.08833 $\pm$ 0.00182 & 2.50 & 0.620 & 0.01580 $\pm$ 0.00033 \\
1.00 & 0.869 & 0.09301 $\pm$ 0.00196 & 2.50 & 0.723 & 0.01617 $\pm$ 0.00034 \\
1.00 & 0.930 & 0.09579 $\pm$ 0.00200 & 2.50 & 0.800 & 0.01619 $\pm$ 0.00032 \\
 \cline{1-3}
2.00 & 0.634 & 0.02527 $\pm$ 0.00053 & 2.50 & 0.846 & 0.01633 $\pm$ 0.00035 \\
2.00 & 0.735 & 0.02606 $\pm$ 0.00053 & 2.50 & 0.949 & 0.01683 $\pm$ 0.00034 \\
2.00 & 0.808 & 0.02652 $\pm$ 0.00053 & 2.50 & 0.963 & 0.01714 $\pm$ 0.00045 \\
 \cline{4-6}
2.00 & 0.877 & 0.02662 $\pm$ 0.00057 & 3.01 & 0.623 & 0.01026 $\pm$ 0.00022 \\
2.00 & 0.938 & 0.02649 $\pm$ 0.00081 & 3.01 & 0.761 & 0.01053 $\pm$ 0.00022 \\
2.00 & 0.953 & 0.02745 $\pm$ 0.00059 & 3.01 & 0.910 & 0.01070 $\pm$ 0.00034 \\
2.00 & 0.963 & 0.02767 $\pm$ 0.00059 & 3.01 & 0.932 & 0.01090 $\pm$ 0.00024 \\
2.00 & 0.968 & 0.02753 $\pm$ 0.00093 & 3.01 & 0.951 & 0.01106 $\pm$ 0.00034 \\
 \cline{1-3}
 \cline{4-6}
\end{tabular}	
\caption{Reduced cross section from Ref.~\cite{Walker:1993vj}.}
\label{tab:walk}
\end{center}
\end{table}\vspace{4mm}
%
%
\begin{figure}[h!]
\begin{center}
\includegraphics[width=\textwidth]{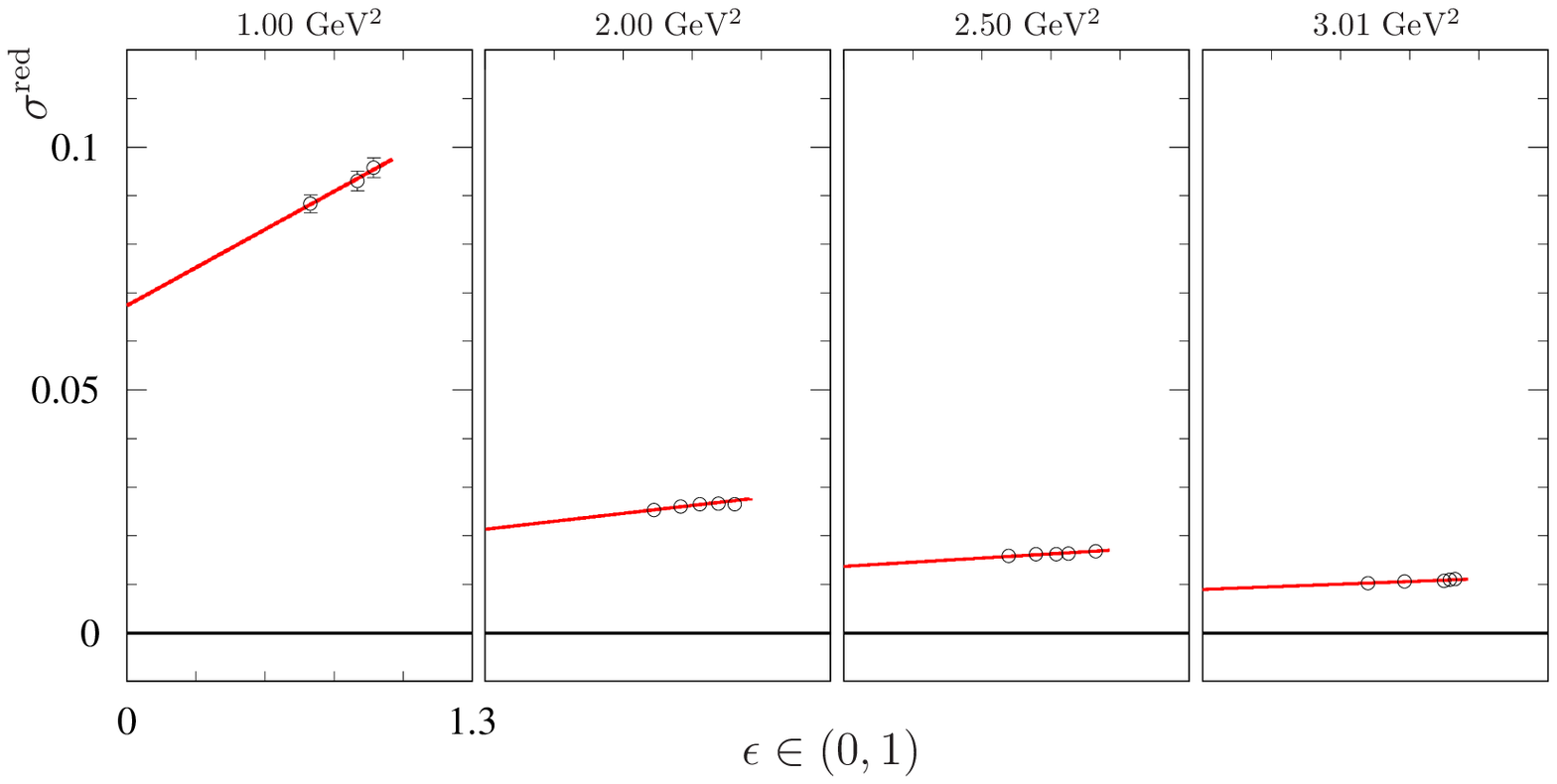}\vspace{-2mm}
\caption{\label{fig:walk}The graphs represent data (open circles) from Ref.~\cite{Walker:1993vj} and fits (red lines) 
of the reduced cross section as a function of $\epsilon$, at the $Q^2$ values reported on the top of each graph.}
\end{center}
\end{figure}
\clearpage
\section{Fit of the data of Ref.~\cite{Litt197040}}
\label{app:litt}
Table~\ref{tab:litt} reports the original data from Ref.~\cite{Litt197040}. Fits to these data, for the five 
$Q^2$ values, are shown in Fig.~\ref{fig:litt}.
%
%
\begin{table}[h]
\begin{center}
	\begin{tabular}{c|c|c||c|c|c}
	$Q^2$  & \multirow{2}{*}{$\epsilon$} & \multirow{2}{*}{$\sigma^{\rm red}$} &
	$Q^2$  & \multirow{2}{*}{$\epsilon$} & \multirow{2}{*}{$\sigma^{\rm red}$}\\
	(GeV$^2$) &&&(GeV$^2$)&&\\
	\hline
	1.00 & 0.955 & 0.09527 $\pm$ 0.00133 & 2.50 & 0.903 & 0.01618 $\pm$ 0.00027 \\
1.00 & 0.932 & 0.09398 $\pm$ 0.00141 & 2.50 & 0.803 & 0.01596 $\pm$ 0.00028 \\
1.00 & 0.918 & 0.09701 $\pm$ 0.00150 & 2.50 & 0.732 & 0.01602 $\pm$ 0.00029 \\
 \cline{1-3}
1.50 & 0.969 & 0.04928 $\pm$ 0.00073 & 2.50 & 0.960 & 0.01665 $\pm$ 0.00023 \\
1.50 & 0.880 & 0.04788 $\pm$ 0.00083 & 2.50 & 0.932 & 0.01692 $\pm$ 0.00034 \\
1.50 & 0.853 & 0.04743 $\pm$ 0.00074 & 2.50 & 0.824 & 0.01639 $\pm$ 0.00029 \\
 \cline{1-3}
2.00 & 0.952 & 0.02779 $\pm$ 0.00047 & 2.50 & 0.733 & 0.01573 $\pm$ 0.00029 \\
2.00 & 0.877 & 0.02651 $\pm$ 0.00043 & 2.50 & 0.672 & 0.01578 $\pm$ 0.00029 \\
 \cline{4-6}
2.00 & 0.814 & 0.02611 $\pm$ 0.00063 & 3.75 & 0.953 & 0.00639 $\pm$ 0.00009 \\
2.00 & 0.772 & 0.02641 $\pm$ 0.00045 & 3.75 & 0.922 & 0.00641 $\pm$ 0.00001 \\
 \cline{1-3}
2.50 & 0.960 & 0.01641 $\pm$ 0.00023 & 3.75 & 0.646 & 0.00601 $\pm$ 0.00012 \\
 \cline{4-6}
\end{tabular}	
\caption{\label{tab:litt}Reduced cross section from Ref.~\cite{Litt197040}.}
\end{center}
\end{table}\vspace{4mm}
%
%
\begin{figure}[h!]
\begin{center}
\includegraphics[width=\textwidth]{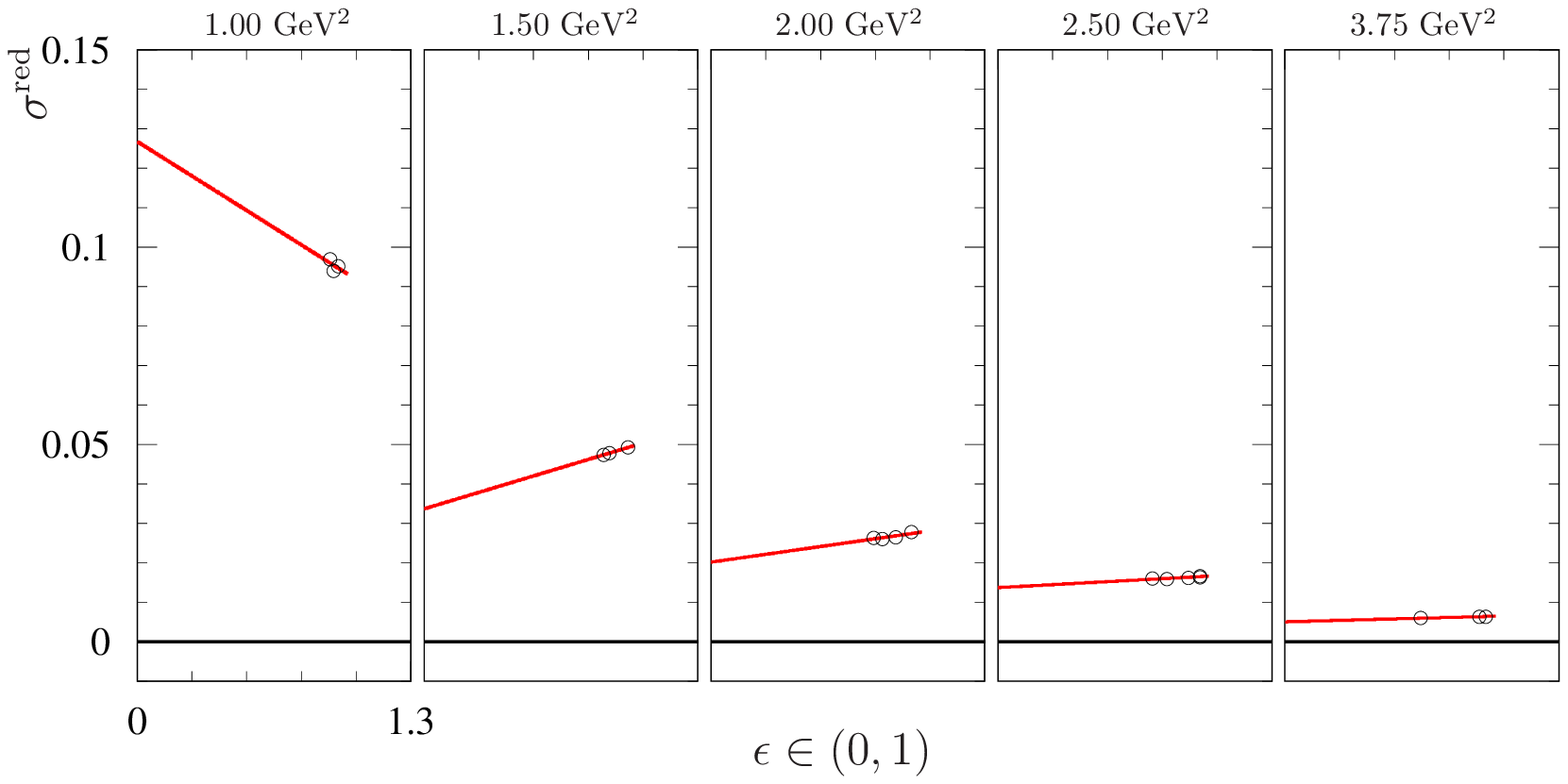}	\vspace{-2mm}
\caption{The graphs represent data (open circles) form Ref.~\cite{Litt197040} and fits (red lines) 
of the reduced cross section as a function of $\epsilon$, at the $Q^2$ values reported on the top of each graph.}\label{fig:litt}
\end{center}
\end{figure}
\clearpage
\section{Fit of the data of Ref.~\cite{Qattan:2004ht}}
\label{app:qatt}
Table~\ref{tab:qatt} reports the original data from Ref.~\cite{Qattan:2004ht}. Fits to these data, for the three 
$Q^2$ values, are shown in Fig.~\ref{fig:qatt}.
%
%
\begin{table}[h]
\begin{center}
	\begin{tabular}{c|c|c||c|c|c}
	$Q^2$  & \multirow{2}{*}{$\epsilon$} & \multirow{2}{*}{$\sigma^{\rm red}$} &
	$Q^2$  & \multirow{2}{*}{$\epsilon$} & \multirow{2}{*}{$\sigma^{\rm red}$}\\
	(GeV$^2$) &&&(GeV$^2$)&&\\
	\hline
	2.64 & 0.117 & 0.01322 $\pm$ 0.00007 & 3.20 & 0.443 & 0.00893 $\pm$ 0.00005 \\
2.64 & 0.355 & 0.01375 $\pm$ 0.00008 & 3.20 & 0.696 & 0.00916 $\pm$ 0.00005 \\
2.64 & 0.597 & 0.01407 $\pm$ 0.00008 & 3.20 & 0.813 & 0.00937 $\pm$ 0.00005 \\
 \cline{4-6}
2.64 & 0.781 & 0.01443 $\pm$ 0.00008 & 4.10 & 0.160 & 0.00466 $\pm$ 0.00003 \\
2.64 & 0.865 & 0.01462 $\pm$ 0.00008 & 4.10 & 0.528 & 0.00490 $\pm$ 0.00003 \\
 \cline{1-3}
3.20 & 0.131 & 0.00858 $\pm$ 0.00005 & 4.10 & 0.709 & 0.00501 $\pm$ 0.00003 \\
 \cline{4-6}
\end{tabular}	
\caption{Reduced cross section from Ref.~\cite{Qattan:2004ht}.}\label{tab:qatt}
\end{center}
\end{table}\vspace{4mm}
%
%
\begin{figure}[h]
\begin{center}
\includegraphics[width=\textwidth]{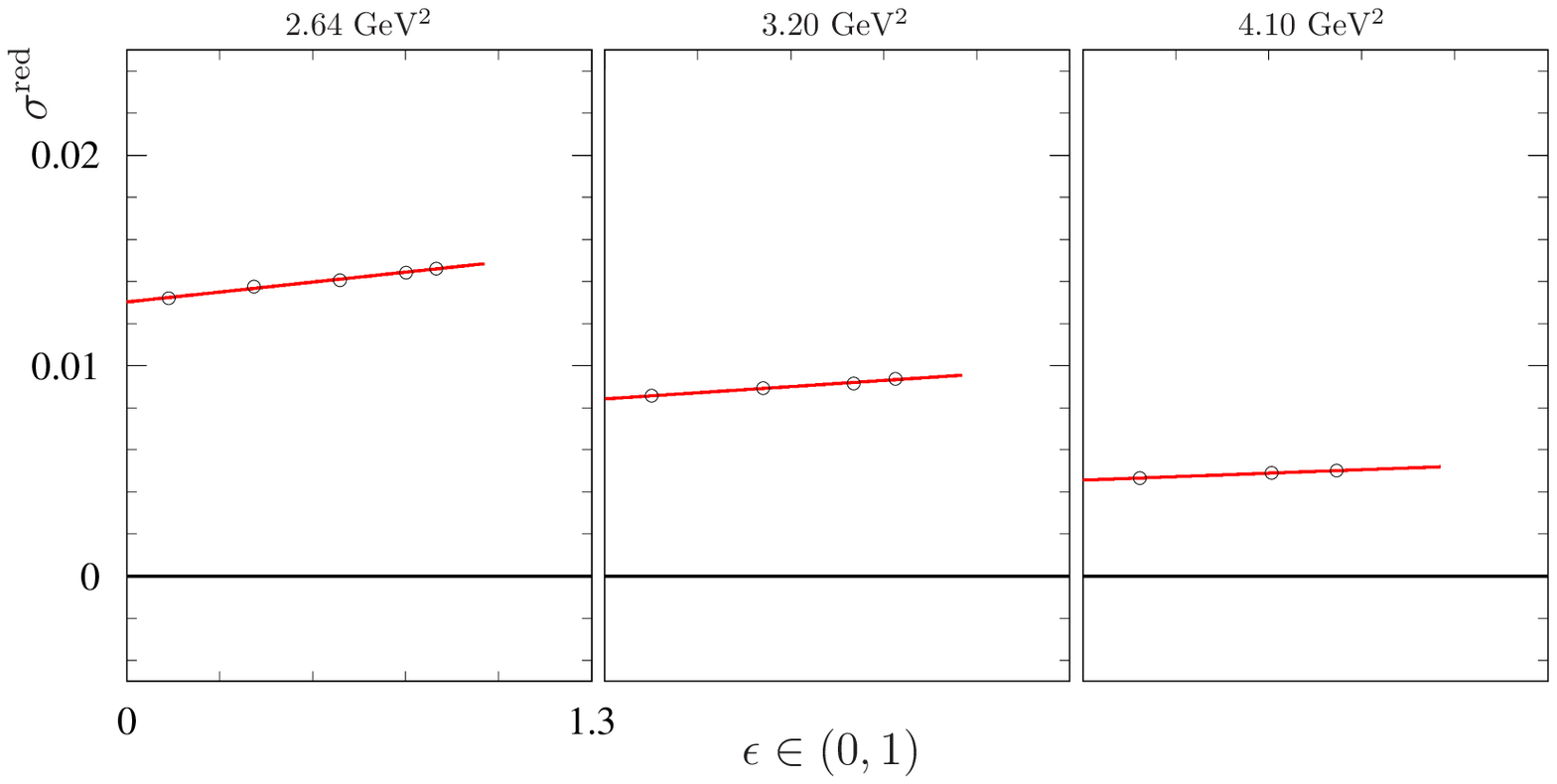}	
\vspace{-2mm}
\caption{\label{fig:qatt}The graphs represent data (open circles) from Ref.~\cite{Qattan:2004ht} and fits (red lines) 
of the reduced cross section as a function of $\epsilon$, at the $Q^2$ values reported on the top of each graph.}
\end{center}
\end{figure}
\end{widetext}


\end{document}